\documentclass[
  amsmath,amssymb,
  preprint,   
]{revtex4-1}

\usepackage[T1]{fontenc}
\usepackage[utf8]{inputenc}

\usepackage{graphicx}
\usepackage{dcolumn}
\usepackage{bm}
\usepackage{amsmath, amssymb}
\usepackage{float}

\usepackage{mathptmx}

\usepackage{xcolor}

\usepackage{caption}
\usepackage{subcaption}

\makeatother
\begin{document}

\preprint{AIP/123-QED}

\title[Measuring FPUT thermalization with Toda integrals]{Measuring FPUT thermalization with Toda integrals}

\author{H. Christodoulidi}
\email{hchristodoulidi@lincoln.ac.uk}  
 \affiliation{University of Lincoln, School of Mathematics and Physics, Brayford Pool Campus, Lincoln, UK}
\author{S. Flach}%
\affiliation{ 
Center for Theoretical Physics of Complex Systems, Institute for Basic Science (IBS), Daenjeon, South Korea
}%

\date{\today}


 
\begin{abstract}
 We assess the ergodic properties of the Fermi--Pasta--Ulam--Tsingou--$\alpha$ model for generic initial conditions using a Toda integral. It serves as an adiabatic invariant for the system and a suitable observable to measure its equilibrium time.  
 Over this timescale, the onset of action diffusion results in ergodic temporal fluctuations.
We compare this timescale with the inverse of the maximum Lyapunov exponent $\lambda$ and its saturation time, which are systematically shorter.
The Toda integral ergodization/equilibrium time is system size independent for long chains, but show dramatic growth when the system size is smaller than a critical one, whose value depends on the energy density. We measure the dependence of energy density on the critical system size and relate this observation to the possible emergence of a  {Kolmogorov--Arnold--Moser (KAM)} regime. We numerically determine the critical energy density of this regime, finding that it approximately decays as $1/N^2$ with the number of particles $N$. 
\end{abstract}

\maketitle

\noindent\textbf{The present work looks at how a classical Hamiltonian model with many degrees of freedom, the so-called Fermi--Pasta--Ulam--Tsingou–$\alpha$ model, approaches thermal equilibrium and explores its overall ergodic behavior under random initial conditions. Our approach differs from typical studies in statistical mechanics, which often focus on the energy spectrum or spectral entropy. Instead, we employ a different set of observables, the integrals of motion of a well-known integrable model, the Toda lattice.
The Toda integrals have proved to be powerful tools for determining the onset of diffusion in phase space trajectories with high precision, modeling the diffusion process, and estimating the equilibrium timescales. Due to their computational simplicity – compared to the energy spectrum – we were able to simulate the system over very long times at very low energies. This allowed us to estimate an energy crossover, below which timescales grow exponentially. We have also compared our findings to the timescales derived from the maximum Lyapunov exponent $\lambda$, where we have concluded that the latter converges faster and reveals weak chaos before phase space diffusion takes place.
}

\section{Introduction}

Modern statistical mechanics aims to understand a variety of complex phenomena, including non-equilibrium behavior, ergodicity breaking, slow thermalization processes and energy localization in physical systems. In this paper, we study   the ergodic properties of the Fermi--Pasta--Ulam--Tsingou--$\alpha$ (FPUT) model \cite{Fermi, Dauxois}, a one-dimensional chain of nonlinearly coupled oscillators, from the point of view of its adiabatic invariants. Our study is part of a broader perspective focusing on the effect of 
adiabatic invariants on non-equilibrium dynamics and the slowing down of thermalization in many-body Hamiltonian systems.
Apart from the celebrated FPUT-$\alpha$ model, this perspective extends to other Hamiltonian systems with known adiabatic invariants, e.g. the Klein--Gordon lattice \cite{Giorgilli12,Giorgilli15} or 
 Mie–Lennard-Jones type potentials in one-dimensional molecular dynamics \cite{Benettin2023}. 

The original FPUT model study was a milestone in the history of nonlinear science. The FPUT problem \cite{book} sparked the birth of ‘experimental mathematics’ \cite{Porter}, as probably the first computational experiment implemented on one of the earliest digital computers. It further contributed to the development of the theory of integrable systems \cite{kdv} through the discovery of solitons in the Korteweg de--Vries equation, and it inspired Chirikov's resonance-overlap theory and the pursuit of a stochasticity threshold \cite{Chirikov73,Chirikov79,Casetti,Shepelyansky}.

This long-standing problem 
has been traditionally studied within the framework of harmonic oscillators and its corresponding harmonic normal modes with nontrivial interaction induced by the nonintegrable nonlinearity addition. Low frequency initial conditions give rise to enduring energy localization patterns.   
This well-visualized perspective offers a clear understanding of the system's distance from its thermal equilibrium, characterized by energy equipartition. A number of important studies report
on measuring the timescales to reach equipartition  \cite{Casetti, Lichtenberg, Benet2009, Benet2011,
danielicampbellflach2017, Parisi,CaratiMaiocchi2012}. 
For weak enough nonlinearity the profiles of the observed enduring energy localization appear in stable domains of the phase space and close to  $q$-breathers, i.e. stable and exact periodic orbits arising from the 
continuation of normal modes \cite{Flach2005,Flach2006,Penati,Karve2024}, or $q$-tori \cite{Christ2010,Christ2013,EPL2017}.
Nevertheless, several theoretical studies agree that such stability regimes vanish toward the thermodynamic limit, with critical energy densities scaling like $\varepsilon_c \propto 1/N^4$, where $N$ is the number of particles. However, this law remains unverified numerically and  contradicted by the original FPUT experiment \cite{Shepelyansky,Bambusi2015}.

Another path of studying  FPUT models adopts a different approach to the ergodicity problem, shifting their focus to its ‘closest’ integrable system, the Toda lattice \cite{Toda}, and its invariants  quantities \cite{Henon, Flaschka}.  Ponno et al.  \cite{Ponno2011} revived the work of Ferguson et al.  \cite{ferguson}, published in 1982, who pointed out the close connection between the two systems.  
Benettin et al.  \cite{Benet2013} provided further evidence on the relevance of FPUT's non-ergodic dynamics being essentially Toda's.  
Goldfriend et al \cite{gold} 
investigated the Lax matrix eigenvalues as observables for the FPUT dynamics, and more
recently Grava et al. \cite{Ponno} proved that Toda’s first integrals constitute a set of adiabatic invariants for the FPUT model. Christodoulidi et al. \cite{CE} studied the first non-trivial integral, which proved to be a promising approach for measuring energy diffusion in the model's phase space and in assessing ergodicity. 

The above discussed approaches focused mainly on the FPUT initial conditions of exciting a few, typically long wavelength, modes. As was shown by Danieli et al \cite{danielicampbellflach2017}, the concepts can be taken to extract timescales of thermalization by studying typical or random initial conditions. When being close to an integrable limit, the conserved actions of the corresponding integrable Hamiltonian are used as observables in the weakly nonintegrable regime. One can then proceed with 
extracting time scales from the time dependent observables as they fluctuate around their Gibbs averages, and also from the evolution of distributions of finite time averages of the actions 
\cite{danielicampbellflach2017}
\cite{Mithun}
\cite{Campbell}.  
It is, therefore, interesting to apply the above concepts to the FPUT model with its known adiabatic Toda invariants, which will be used as observables instead of the harmonic mode actions of the corresponding linear integrable limit.

Another set of timescales which characterize thermalization can be obtained from the computation of the entire Lyapunov spectrum \cite{benettin1980lyapunov,skokos_lyapunov_2010}. Inverting the exponents yields a Lyapunov time spectrum. Weakly nonintegrable perturbations span a nonlinear interaction network among the (countable) actions of an integrable Hamiltonian. This network is typically either long-ranged or short-ranged \cite{danielicampbellflach2017,danieli_dynamical_2019,mithun_dynamical_2019,mithun_fragile_2021}. The FPUT model falls into the long-ranged network class. Most likely this is also true for the Toda lattice, if perturbed slightly such that to arrive at a nonintegrable FPUT partner model. The study of the scaling of the Lyapunov spectrum upon approaching an integrable limit \cite{malishava_lyapunov_2022,malishava_thermalization_2022,malishava_thermalization_2022-1,lando2023thermalization,zhang2024thermalization,zhang2025observation} revealed that Lyapunov spectra in long-range networks scale uniformly, i.e. the entire spectrum scales with say the largest Lyapunov exponent. Moreover, this appears to be also true when comparing to ergodization time scales of action observables of the corresponding integrable limit - while the absolute values may all vary from observable to observable, they all scale similar to the inverse of the largest Lyapunov exponent, which we hereafter coin $T_{\Lambda}$. This apparently homogeneous and uniform thermalization time scaling appears to be universal for long-range networks. We will therefore also compute $T_{\Lambda}$ and compare its absolute value and scaling properties to ergodization timescales. The above-mentioned short range network instead falls into a different universality class, which we will not further discuss and consider here. 

This paper is structured as follows: In section \ref{intro}, we introduce the FPUT and Toda models, the first non-trivial Toda integral and its basic features. In section \ref{ch2}, we demonstrate the superiority of the Toda integral as an adiabatic invariant observable for the FPUT model, by comparing its ergodization with the energy spectra and the spectral entropy in the original FPUT experiment of exiting the first normal mode for $N=31$ particles. In section  \ref{ch3},\ref{chapter_diffusion} we numerically compute and analyze the evolution of the Toda integral starting from random initial conditions, and fit it with a sigmoid diffusion law. We examine the system's two main timescales -- the onset of diffusion and the equilibrium time -- and determine their power-law dependence on FPUT's energy density. In section \ref{chLE}, we compare these timescales with the inverse of the Lyapunov exponent and its saturation time. Finally, in section \ref{final} we evaluate the critical energy density, which distinguishes exponentially long timescales from uniform, and investigate its asymptotic behavior in the thermodynamic limit.


\section{FPUT and Toda models}\label{intro}

The FPUT-$\alpha$ model lattice model consists of $N$ particles coupled only with their nearest neighbors, representing the simplest one-dimensional analog of atoms in a crystal. We denote the displacement of the $n$--th particle  from its equilibrium position by $q_n$  and  the corresponding canonically conjugate momentum by $p_n$. Fixed boundary conditions are set by considering $q_0=q_{N+1}=p_0=p_{N+1}=0$. For simplicity, throughout this paper, we set the nonlinearity strength $\alpha=1$  and refer to the FPUT-$\alpha$ model simply as the FPUT model. The relevant control parameter which tunes the effective nonlinearity strength is then the energy density \footnote{The relevant perturbation parameter in FPUT-$\alpha$ model is $\alpha \sqrt{\varepsilon}$, hence we fix $\alpha=1$ and modify only the energy density $\varepsilon$.} $$\varepsilon = \frac{E}{N}~~,$$
i.e. the total conserved energy of the system divided by the number of particles $N$.
The dynamics of the FPUT model is described by the Hamiltonian
\begin{eqnarray}\label{fpuham} 
H_{FPUT}=\sum_{n=0}^{N} \big[ {p_n^2 \over 2}  
+ {(\delta q_n)^2 \over 2}  + { (\delta q_n)^3 \over 3}  \big]
\end{eqnarray}
where $\delta q_n$ is a short notation for the relative displacement terms $q_{n+1}-q_n$.  
The quadratic part of the above Hamiltonian corresponds to the harmonic oscillator chain with mode energies
\begin{eqnarray}\label{energyspectrum} 
E_k= {P_k^2+ \omega_k ^2 Q_k^2\over 2}
\end{eqnarray}
where
\begin{eqnarray}\label{can} 
\begin{pmatrix}
Q_k \\
P_k 
\end{pmatrix}
= 
\sqrt{2\over N+1}\sum_{n=1}^{N} \begin{pmatrix}
q_n \\
p_n 
\end{pmatrix}
 \sin\left({n k\pi\over N+1}\right)
 \end{eqnarray}
are the mode coordinates and momenta, respectively and  $\omega _k = 2 \sin( \frac{k \pi}{2(N+1)} ) $, $k=1,\ldots,N$ 
the harmonic frequencies.

\begin{figure}
\centering
    \includegraphics[width=0.75\linewidth]{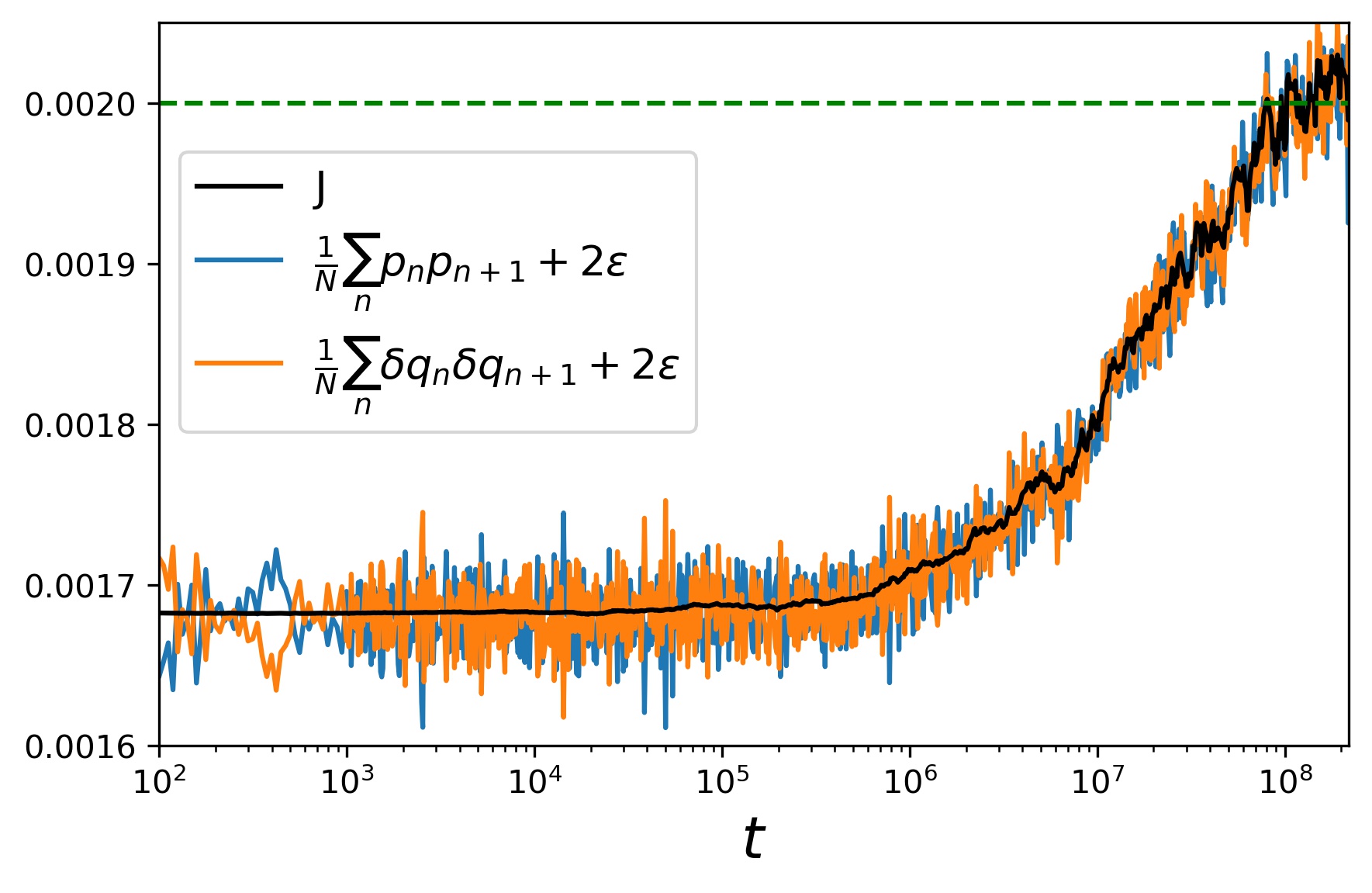 }
    \caption{ The Toda integral $J$ from (\ref{j1}) along a trajectory of the FPUT model with random initial conditions, together with the sum of momenta products $p_n p_{n+1}$ (blue) and the sum of relative displacement differences $\delta q_n \delta q_{n+1}$ (orange). The system parameters are $N=2047$ particles at the energy level $\varepsilon = 10^{-3}$ for a single random initial condition without time-averages. The (green) dashed line marks  the
Gibbs average value $J_{eq}$ reached at approximately $t=10^8$. }
    \label{JandQT}
\end{figure}

The FPUT Hamiltonian (\ref{fpuham}) is the third--order truncation of Toda Hamiltonian 
\begin{eqnarray}  \label{B1a}
H_{T}=  \sum_{n=0}^{N} \big[ {p_{n}^{2} \over 2}  +  
\frac{ e^{2 \delta q_n }  -1}{4 } \big], 
\end{eqnarray}
whose dynamics is completely integrable. This model possesses $N$ independent integrals of motion, i.e. $N$ first integrals in involution with respect to the standard Poisson bracket.  

\begin{figure}
\centering
    \includegraphics[width=0.75\linewidth]{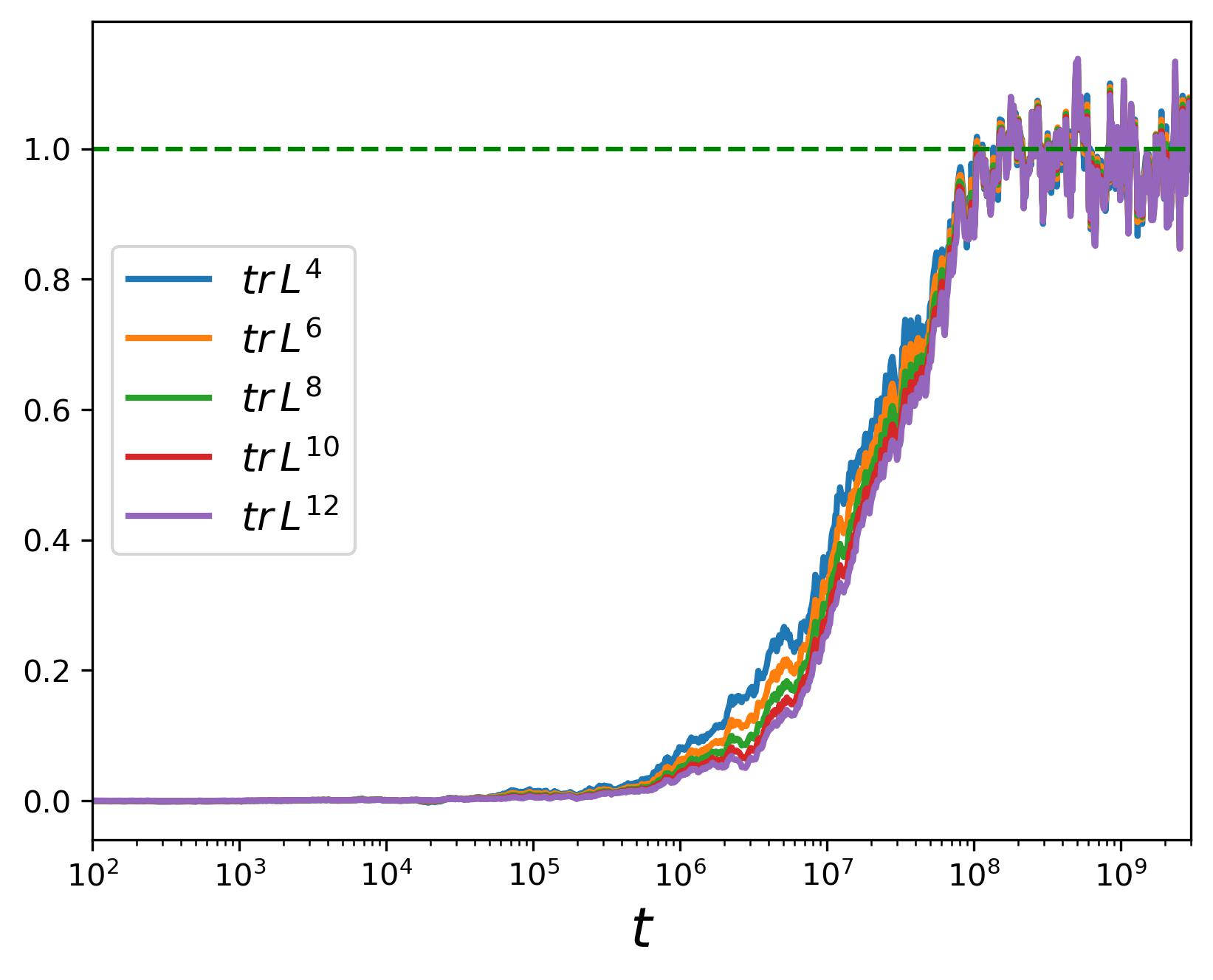 }
    \caption{  { Some first Toda integrals along the FPUT dynamics are derived from the traces of even $L$--matrix powers \cite{Benet2013}. The chosen initial condition is the same as in Fig.\ref{JandQT}, i.e.  $N=2047$, $\varepsilon = 10^{-3}$ for a single random initial condition.
    These integrals have been rescaled to start from value zero and reach equilibrium at value one. }  }
    \label{Traces}
\end{figure}
In this paper we use the Toda integral 
\begin{eqnarray}\label{j1}
J={1 \over 2 N} \sum _{n=0}^{N}  \big[  p_n^4    +  
 e^{2 \delta q_n } (p_n^2 + p_n p_{n+1}+ p_{n+1}^2)   
+\nonumber\\
\frac{e^{2 \delta q_n }}{8  }  ( e^{2 \delta q_{n-1} }+e^{2 \delta q_n } +e^{2 \delta q_{n+1} }) 
- \frac{3} { 8 } \big]  
\end{eqnarray}
derived from the trace of the fourth power of the $L$--matrix \cite{Flaschka,Henon,Benet2013} 
and  {modified the original integral by subtracting the constant $3/8$ and dividing by $N$, in order to obtain an intensive quantity that does not depend on the system size. $J$'s independence on $N$ becomes evident from its quadratic approximation given below.}
This integral was also studied in Refs. [\onlinecite{Ponno2011},\onlinecite{CE}].

To quantify ergodization and thermalization in the FPUT problem, we define
 $T_{eq}$  as the equilibrium time when $J$ reaches its Gibbs value $ J_{eq} $ for the first time. From (\ref{j1}) we find that $$ J_{eq} \approx 2 \varepsilon $$ in leading order for small $\epsilon$. In particular, Taylor expansions of the exponential terms in (\ref{j1}) yield the quadratic approximation of $J$
\begin{eqnarray}\label{quad}
J  \approx 2\varepsilon +  \frac{1}{2N} \sum_n ( p_n p_{n+1} + \delta q_n \delta q_{n+1} ) 
~~ .
\end{eqnarray}

At low energies in the FPUT system, $J$ in (\ref{j1}) and its quadratic approximation (\ref{quad}) are almost indistinguishable. As shown in Fig.\ref{JandQT}, the two sums in the above expression are the essential components of the diffusion mechanism in the FPUT model. They 
depend on the nearest-neighbor momentum and relative position couplings (covariance), and complement each other in producing the adiabatic invariant $J$. 
That is, each of these sums oscillates in opposite phase to the other, and together they combine the smooth curve  (\ref{quad})  that absorbs these oscillations. As the system approaches equilibrium, momenta $\mathbf{p}$ and relative positions $\delta \mathbf{q}$ become uncorrelated,  {which leads to 
$\langle p_n p_{n+1} \rangle =  \langle \delta q_n  \delta q_{n+1} \rangle = 0$, and thereby shows that 
$J_{eq}$ is approximately equal to $2 \varepsilon$}.
In \cite{Parisi}, Parisi employed a similar indicator to measure ergodization; the time-averages of the normalized sum of nearest-neighbor position correlations $\sum _n  q_n q_{n+1} / \sum _n q_n^2$. 


It is worth mentioning that the first Toda integral, the Hamiltonian of the Toda system (\ref{B1a}), does not contain a mechanism similar to  (\ref{quad}) and remains quasi-constant regardless of FPUT's dynamical stage. Its Gibbs average value is approximately $\langle H_{T} \rangle = E$, hence no deviations are expected.  

The other Toda integrals, however,  { display a dynamical behavior that is more similar to $J$. Fig.\ref{Traces} displays a few Toda integrals, derived by $J_k=\mathrm{tr}L^{k}$, $k=4,6,8,10,12$ and rescaled to start from $\Tilde{J}_k(0)=0$ and reach equilibrium at 1, i.e. $$\Tilde{J}_k(t)=\frac{J_k(t)-J_k(0)}{J_k(t_f)-J_k(0)} \quad .$$
All these first integrals exhibit a quite similar diffusive behavior and share the same Gibbs equilibrium timescale. By carefully inspecting their differences, we find  that $\Tilde{J}_4$ rises more quickly as compared to the other integrals (Fig.\ref{Traces}). 
We therefore think that the choice of $k=4$ Toda integral (\ref{j1}) is both computationally simpler, and slightly more sensitive to phase-space diffusion, hence the most reliable choice among them. }

\section{Superiority of Toda integrals}\label{ch2}
\begin{figure}
    \centering
    \includegraphics[width=0.75\linewidth]{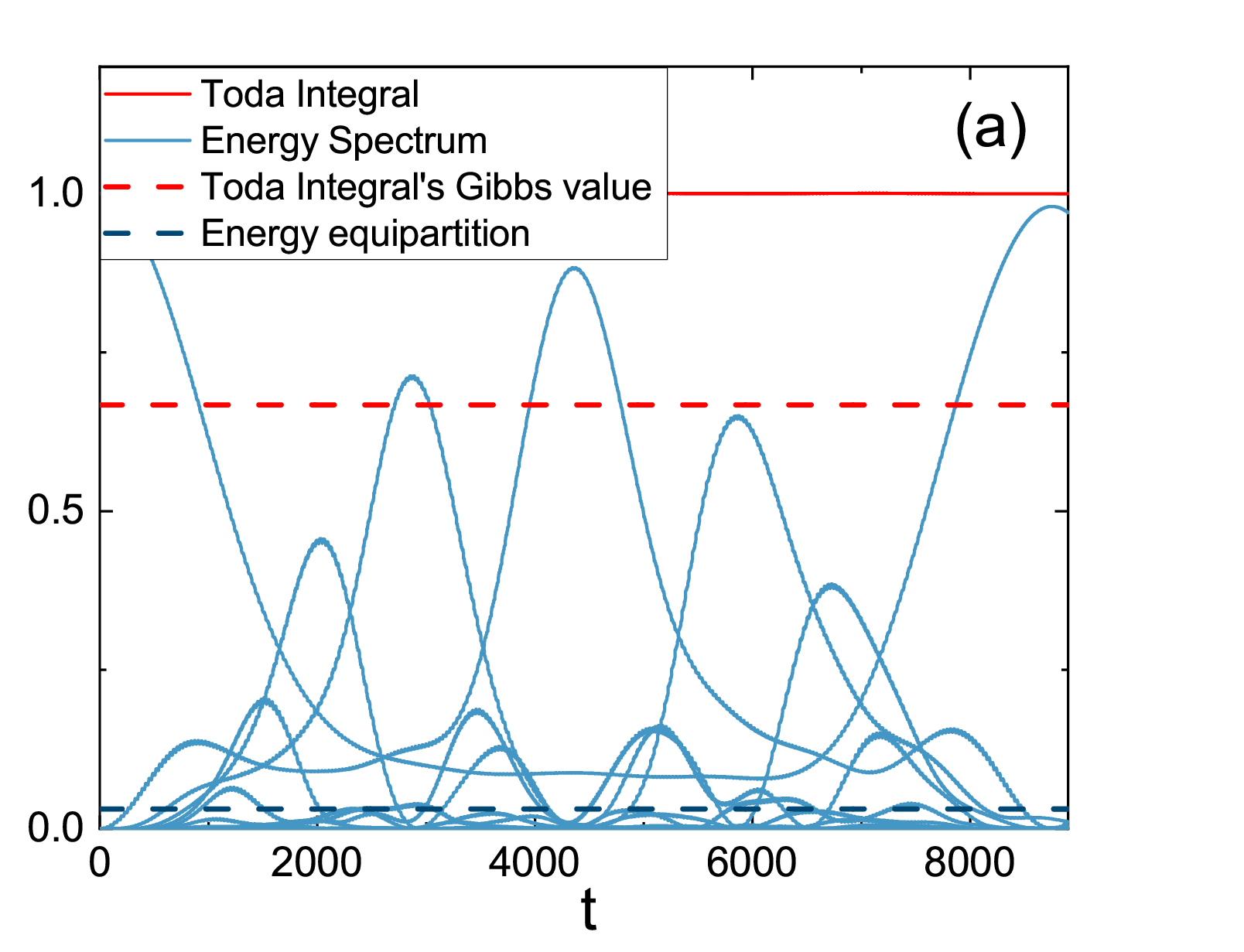}
    \includegraphics[width=0.75\linewidth]{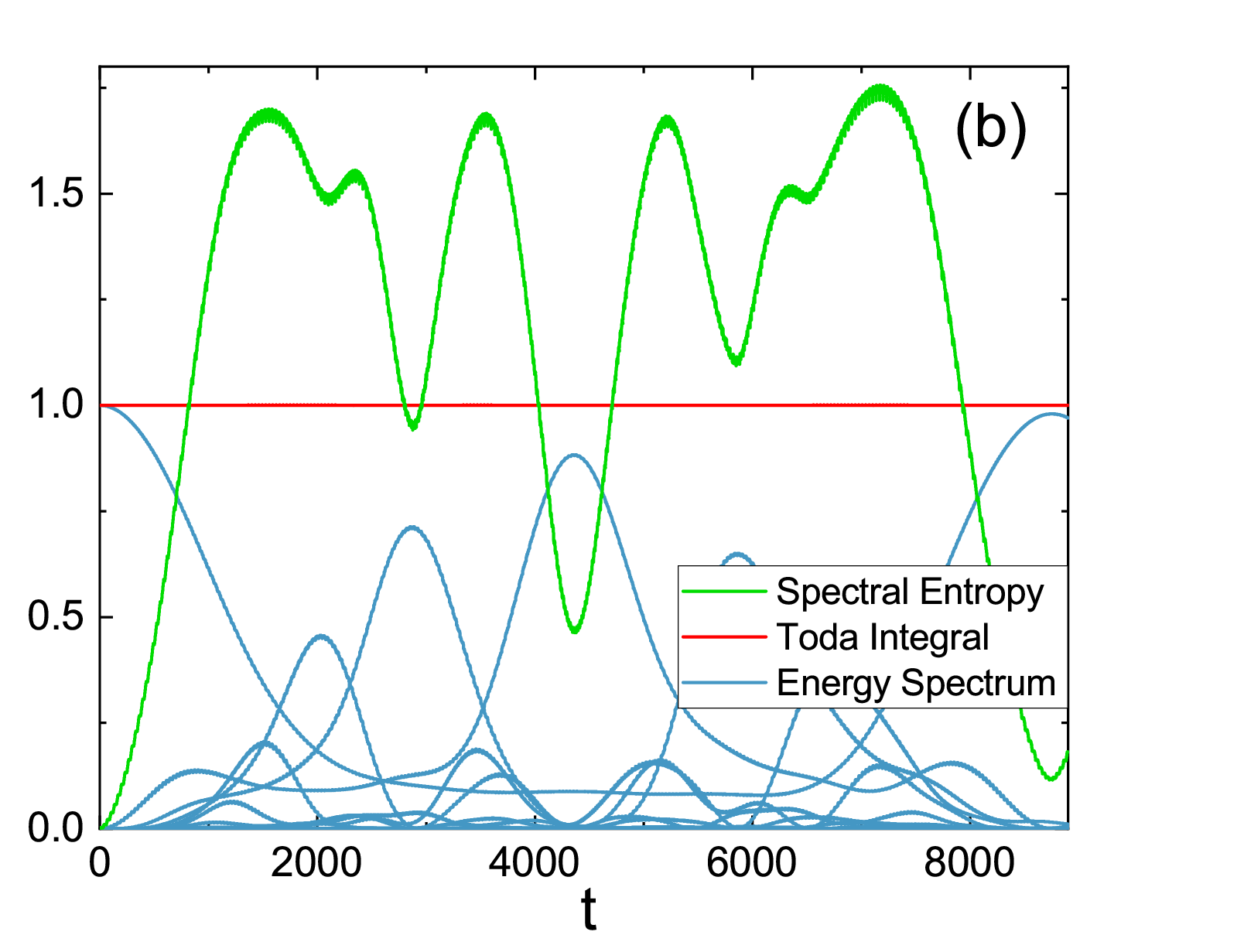}
    \caption{Excitation of the first normal mode in the original FPUT experiment. 
    (a) The blue curves represent the 
    normalized energy spectra $\varepsilon _k$, and the solid red line at 1 the normalized Toda integral $J(t)/J_0$ from (\ref{j1}). The two dashed lines about 0.03 with blue and 0.65 with red  
    represent the Gibbs average value for the energy spectra $\varepsilon_k(t)$, and the Toda integral $J(t)/J_0$, respectively.
    (b) The plot contains the normalized energy spectra and the normalized Toda integral from panel (a), contrasted with the spectral entropy $S(t)$ in (\ref{spectral}). The Gibbs value for $S(t)$ is at 3.47, which has not been included in this plot.
    }
    \label{fig:Eks}
\end{figure}

FPUT models have typically been studied within the framework of harmonic energy spectra $E_k, k=1,\ldots,N$ (\ref{energyspectrum}), as well as in terms of the spectral entropy. By $$\varepsilon_k = \frac{E_k}{E}$$ we denote the normalized harmonic energies, $E$ being the total energy of the system.  The 
spectral entropy reads
\begin{eqnarray}\label{spectral}
S = - \sum_{k} \varepsilon_k \ln \varepsilon_k ~~ .
\end{eqnarray}
The equi-distribution of energy among the harmonic normal modes (energy equipartition) indicates that the system has reached thermodynamic equilibrium, characterizing it as ergodic. 
Fermi, Pasta, and Ulam \cite{Fermi}, together with Tsingou \cite{Dauxois}, used the excitation of the first normal mode as an initial condition to observe this equipartition. Their choice for $N=31$ particles at low energies resulted in  the so-called FPUT-recurrences of the energy spectra, shown in Fig.\ref{fig:Eks}(a), that display quasi-periodic behavior instead of ergodic.

In this paper, we regard a different approach to this problem. We consider a constant of motion for the Toda model,  $J$, as expressed in (\ref{j1}). The main reason for choosing a Toda-based variable is that $J$ accurately identifies energy diffusion, in the sense that it 
is flat (constant) as long as there is no action diffusion among it and the other Toda integrals.
Indeed, as is evident from the 
original FPUT experiment, both panels of Fig.\ref{fig:Eks} display the evolution of the energy spectra, as well as the normalized Toda integral. The latter does not fluctuate while the system exhibits quasiperiodic oscillations of mode energies. In panel (b), the spectral entropy $S(t)$ from Eq.(\ref{spectral}) is also included. Both the energy spectra and the spectral entropy show recurrent behavior. The Toda integral, on the other hand, remains constant in time absorbing inessential recurrent oscillations.

Comparing each of the observables with their Gibbs average value is crucial for evaluating the size of  `real' fluctuations. As an example, the individual normal mode energies
in Fig.\ref{fig:Eks}(a) fluctuate and intersect their Gibbs average value at $1/N$ (blue dashed line around 0.03). The spectral entropy does not intersect its Gibbs average value, yet fluctuates strongly with amplitudes of the order of its thermal mean. The normalized Toda integral (red line at 1) is simply constant, and does not intersect its Gibbs average value (dashed line about  $J_{eq}/J_0 = 0.65$), maintaining a bounded distance from $J_{eq}/J_0$ throughout the simulation time. In Fig.\ref{fig:Eks}(b), the energy spectra and the Toda integral are compared to the spectral entropy. The latter fluctuates with at  $70\%$ higher amplitude compared to the sum of the normalized energy spectra $\sum_k \varepsilon_ k \approx 1$, though it does not intersect its Gibbs value, which is around $S_{eq}=3.47$.

\section{Evolution of the Toda integral along generic FPUT dynamics}\label{ch3}
The numerical results presented in this section and subsequent ones  are based on random initial conditions. We choose random positions and momenta extracted from uniform distributions, which are then rescaled with a Newton--Raphson method to attain the precise energy target value $E$.   {An illustration of the energy spectrum for this type of initial condition is shown in Fig.\ref{Eks}. }

Instead of usual running time-averages we perform averages of the Toda integral over a few realizations. In this way, we observe a convergence at equilibrium at earlier times, while having control over data spreading. In the following figures, each curve representing the evolution of $J$ along the FPUT dynamics is averaged over 20 realizations, producing a smoother and easier to interpret curve.

\begin{figure}
    \centering
\includegraphics[width=0.75\linewidth]{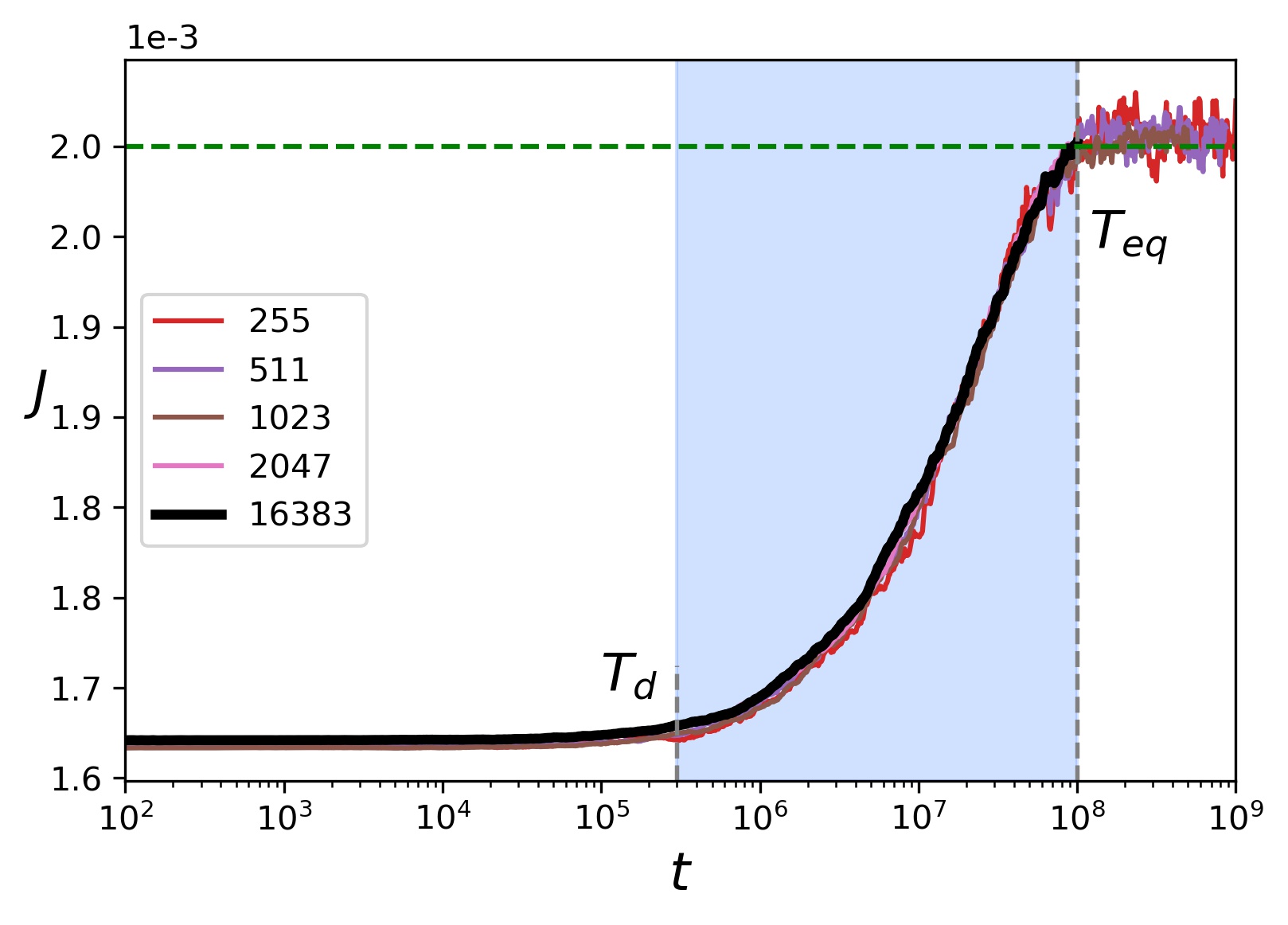}
    \caption{The evolution of the Toda integral at energy density  $\varepsilon = 10^{-3}$ and for different system sizes $N=255, 511,1023, 2047, 16383$.  $J$-curves correspond to averages derived from 20 realizations. 
    The light-blue area is the energy diffusion window which marks two timescales: the onset of diffusion $T_d=2 \times 10^5$, and the equilibrium time $T_{eq} =10^8$, where $J$ reaches its Gibbs value $J_{eq} \approx 2 \varepsilon$.
    The coefficients of variation (ratio of the standard deviation to the mean) are $2.4 \%$, $1.8\%$ $1.5 \%$ $0.9 \%$ and $0.03 \%$ respectively  {, as shown in Fig.\ref{std1} in Appendix}. 
    }
    \label{paradigm}
\end{figure}

A representative example of the evolution of the Toda integral (\ref{j1}) for this class of initial conditions is shown in Fig. \ref{paradigm}, where the route to thermalization is independent of the system size. 
By keeping the energy density $\varepsilon$ constant and considering different system sizes, such as $N=255, 511,1023, 2047, 16383$, the resulting curves show some universality. 
Furthermore, this figure indicates the existence of two timescales. The shaded area in the plot highlights the time-frame where $J$ increases by detaching from its initial value $J(0) \approx 1.67 \times 10^{-3}$ at $t=2 \times 10^5$ until eventually reaching its Gibbs value $J_{eq} \approx 2 \varepsilon = 2 \times 10^{-3}$ at $t = 10^8$.  

The first timescale marks the onset of action diffusion in the system, and will be called $T_d$ (diffusion time), while the second timescale is the equilibrium time at which $J$ reaches its Gibbs value for the first time, which will be denoted by $T_{eq}$. Both timescales in this example are independent of $N$.

\section{ Energy diffusion and equilibration laws}\label{chapter_diffusion}

In this section our focus is on quantifying and analyzing the energy diffusion of the FPUT model. To our best knowledge, such a study remains under-explored and has not received the attention it merits. We aim to provide a clear law modeling the energy drift in phase space, as it is illustrated in Fig.  \ref{paradigm}. Such a law can approximate or predict: the rate of diffusion, the diffusion time $T_d$, as well as the  equilibrium time $T_{eq}$.
From the graphs, the diffusion law is a sigmoid curve (in logarithmic time) where
\[
J(t, \varepsilon) = 
\begin{cases} 
    J_0 & \text{for } t \leq T_d \\
   {(J_0+J_{eq})}/{2} & \text{at } t = t_0, \text{ (midway to equilibration) }  \\
    J_{eq} & \text{for } t \geq T_{eq}
\end{cases}
\]
A sigmoid curve satisfying the above conditions and describing energy diffusion in the phase space of the FPUT model is
\begin{eqnarray}\label{sigmoid}
J(t, \varepsilon) = J_{eq} - 
\frac{ J_{eq} - J_0 }{1 + (t/t_0)^{\gamma( \varepsilon)}} ~~~.
\end{eqnarray}
As shown in Fig. \ref{sigmoid_fittings} in Appendix, the evolution of the Toda variable $J$ along the FPUT dynamics is well-predicted by  Eq.(\ref{sigmoid}) for a range of $\gamma$ values   {that depend on the energy $\varepsilon$. In particular, 
we perform a numerical fit of Eq.(\ref{sigmoid}) to the data, where the exponent $\gamma(\varepsilon)$ is determined by a best-fit algorithm. It}
characterizes the system as diffusive ($\gamma \approx 1$) at higher energy densities and subdiffusive at lower energies, where $\gamma <1$, with a threshold around $10^{-3}$. This is due to the deviation of $J$ being approximately 
 $$\Delta J (t) = J(t)-J(0) 
 \simeq    \frac{ (J_{eq}-J_0)  }{t_0^{\gamma}} \cdot  t^{\gamma}  \propto c \cdot t^{\gamma} ~~,$$
at times around $T_d$, where $J$ detaches from its initial state.

\begin{figure}
    \centering
    \includegraphics[width=0.75\textwidth] {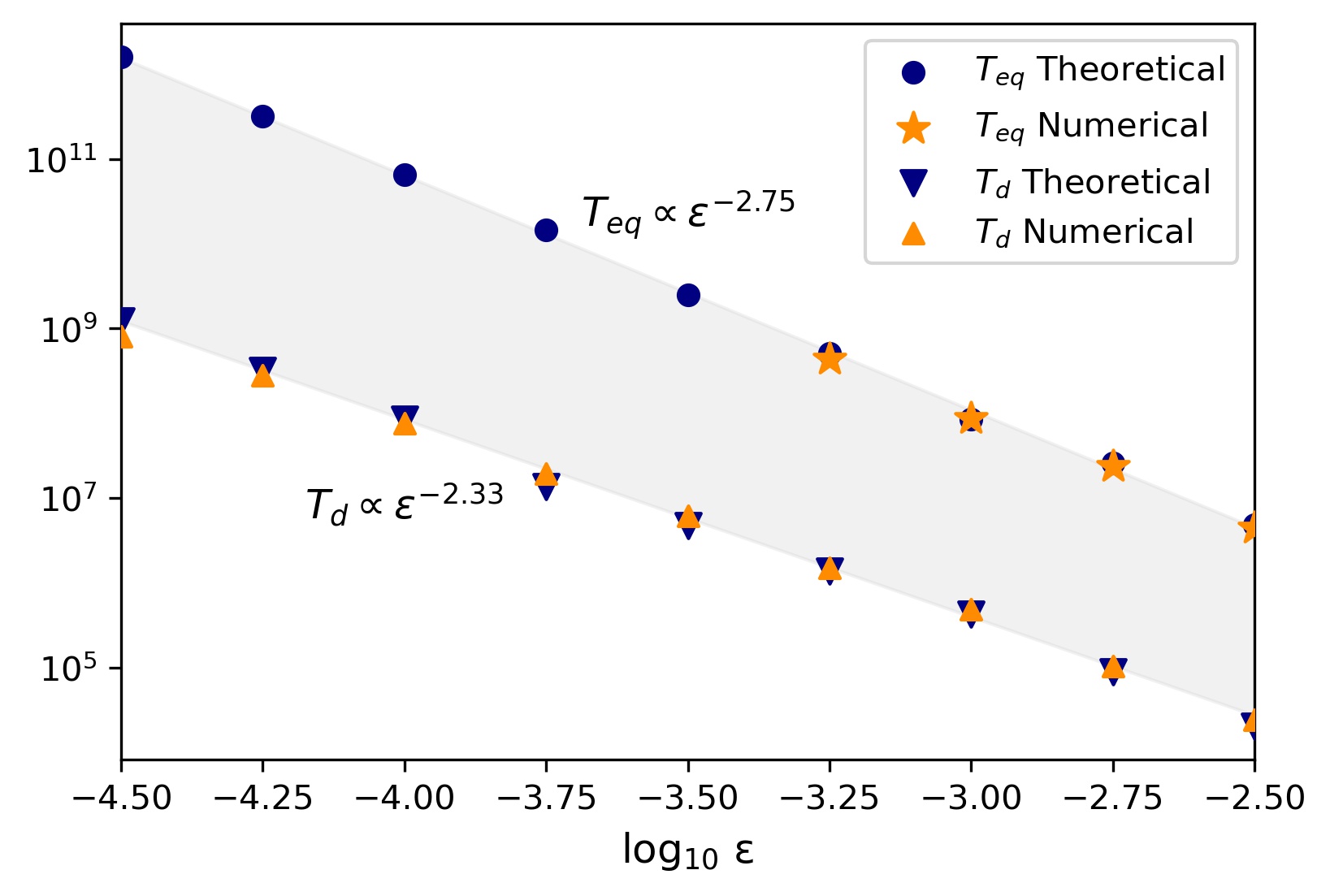}
\caption{ Timescales versus the logarithm of the energy density $\varepsilon$. The numerical timescales for $T_d$ and $T_{eq}$ are in orange, and their theoretical prediction derived from Eq.(\ref{sigmoid}) in blue.
The gap between $T_d$ and $T_{eq}$ increases by lowering the energy density.  At very low energies, $T_d$ may remain observable, whereas $T_{eq}$ is only predicted by Eqs.(\ref{sigmoid},\ref{teq}).   } \label{sigmoid_timescales}
\end{figure}

The sigmoid diffusion law (\ref{sigmoid})  provides an effective theoretical prediction for the equilibrium time $T_{eq}$. 
This law aligns well with the numerical data shown in Fig.\ref{sigmoid_timescales}. 
In  cases of ultra-low energies, such as those shown in Fig.\ref{sigmoid_fittings}(e) and (f) in Appendix, numerical timescale could not be obtained, since the computations require very long times to finalize. However, under (\ref{sigmoid}),  an estimate of the equilibration time can be determined, even for an incomplete curve.
These estimates  are 
\begin{eqnarray}\label{td}
T_d (\varepsilon) \approx  0.04 \varepsilon ^{-2.33}
\end{eqnarray}
for the diffusion timescale and 
\begin{eqnarray}\label{teq}
T_{eq} (\varepsilon) \approx 0.54 \varepsilon ^{-2.75}
\end{eqnarray}
for the equilibrium timescale, shown in Fig.\ref{sigmoid_timescales} together with the numerical results for $T_d$ and $T_{eq}$ for comparison. 
It is noteworthy that a similar
study for the mid-timescale $t_0$, for which $J(t_0)=(J_0+J_{eq})/{2}$, yields $t_0(\varepsilon) \propto \varepsilon ^{-2.67}$, which is closer to $T_{eq}$ timescale (\ref{teq}).

The difference between the exponent values $-2.33$ and $-2.75$
in these two timescales 
 shows that in the 
lower energy limit the gap between $T_d$ and $T_{eq}$ enlarges. This is in agreement with the diffusion exponent $\gamma(\varepsilon) \propto \varepsilon ^ {0.15}$ (Appendix \ref{AppSigm}) and the reduction in the steepness of the sigmoid curves.  
Since the laws (\ref{td}) and (\ref{teq})  represent {\it intensive}
quantities, they are expected to provide a good approximation for any
system above $N \geq 2047$ degrees of freedom. 

  {Comparing these timescale exponents with literature findings, in Ref. \onlinecite{Benet2011} the authors report an equipartition law with an exponent around $-2.25$ for the $\alpha + \beta$ model, where $\alpha =1$ and $\beta$ ranges from $4/3$  to $4$. When $\beta$ approaches Toda's Taylor expansion, $\beta_{T}=2 \alpha^2/ 3$, then these timescales show a notable rise  \cite{Benet2013}.} 
 It is worth noting that their analysis is based on a different set of initial conditions; the excitation of the lowest $10 \%$ of normal modes with random phases, while their observations focus on $t_0$ timescale of the tail-energy $\eta$, defined as the sum of energies in the last half of the normal modes. The tail-energy $\eta$ approach \cite{Benet2011} exhibits a sigmoid behavior similar to $J$.  

\section{ Comparison with Lyapunov exponents}\label{chLE}
To obtain the maximum Lyapunov exponent $\lambda$, it is necessary to numerically compute the finite-time maximum Lyapunov exponent 
\begin{equation}
\lambda(t) = \frac{1}{t} \ln \frac{\| \mathbf{w}(t) \|}{\| \mathbf{w}(0) \|},
\end{equation}
where $\mathbf{w}(t)$ is a vector on the tangent space of the phase space determined by
$$ \dot{\mathbf{w}} = \left[ \mathbf{J}_{2N} \cdot \mathbf{D^2_H}(\mathbf{\textbf{q,p}}) \right] \cdot \mathbf{w}.
$$
In the latter system, $\mathbf{J}_{2N}$ is the standard symplectic matrix and $\mathbf{D^2_H}$ is the Hessian matrix of the FPUT Hamiltonian. The maximum Lyapunov exponent $\lambda$ is defined as the limit $\lim_{t \to \infty} \lambda(t)$. 

In practice, $\lambda(t)$ initially decays as a power-law in time, then slows down and, finally, converges to a positive value $\lambda$ at a finite time. 
At this point, we terminate the numerical simulation, and this value is taken as the  maximum Lyapunov exponent of the given trajectory. 
We define the saturation time $T_{\Lambda S}$ of the maximum Lyapunov exponent as the time at which the power-law
decay reaches the value of $\lambda$. This choice has been preferred to the numerical convergence time of $\lambda$, since the latter gives significant variations. For example, some  $\lambda(t)$ curves fluctuate for a longer period until they numerically convergence to $\lambda$. 


\begin{figure}
    \centering
\includegraphics[width=0.75\textwidth]{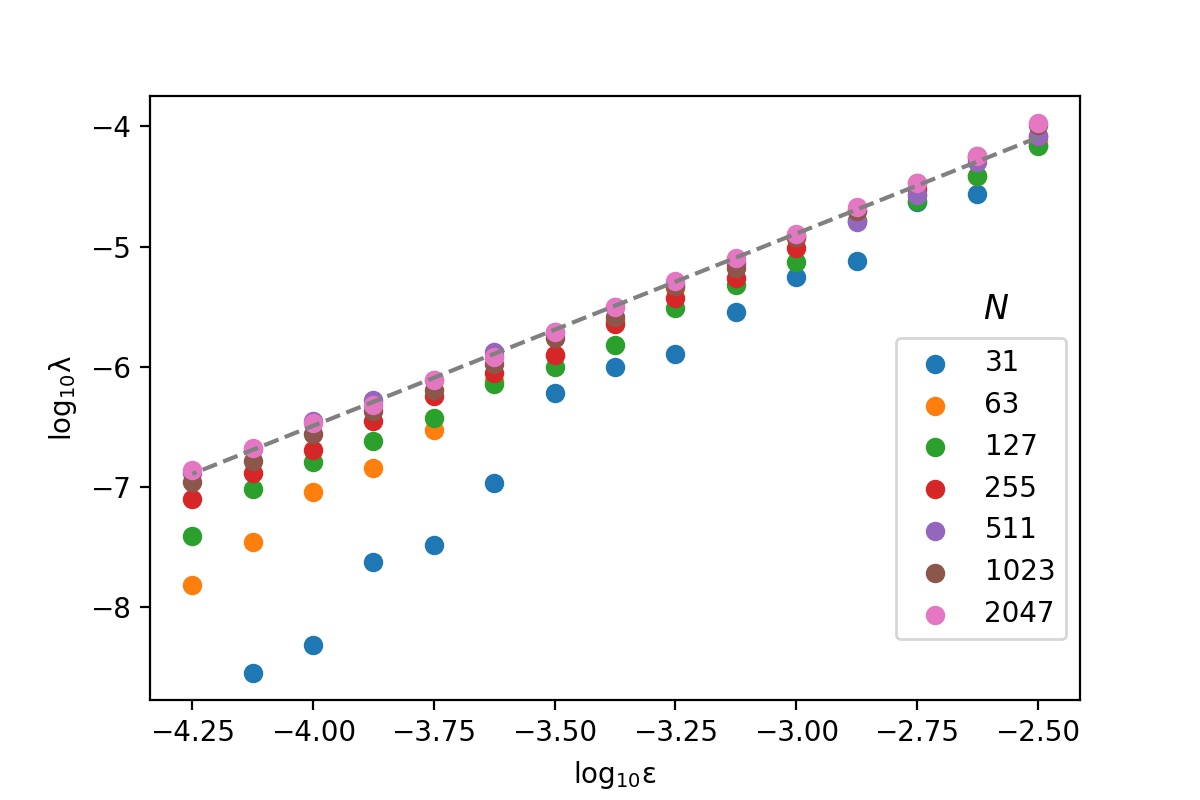}
    \caption{Maximum Lyapunov exponents plotted versus the energy density for various system sizes. Good agreement for $N$ ranging from $511$ to $2047$, fitted by the dashed line with slope $1.6$. Evident disagreement for $N=31$ and 63 particles.}    
     \label{All_LEs}
\end{figure}

  \begin{figure}
    \centering
     \includegraphics[width=0.75\linewidth]{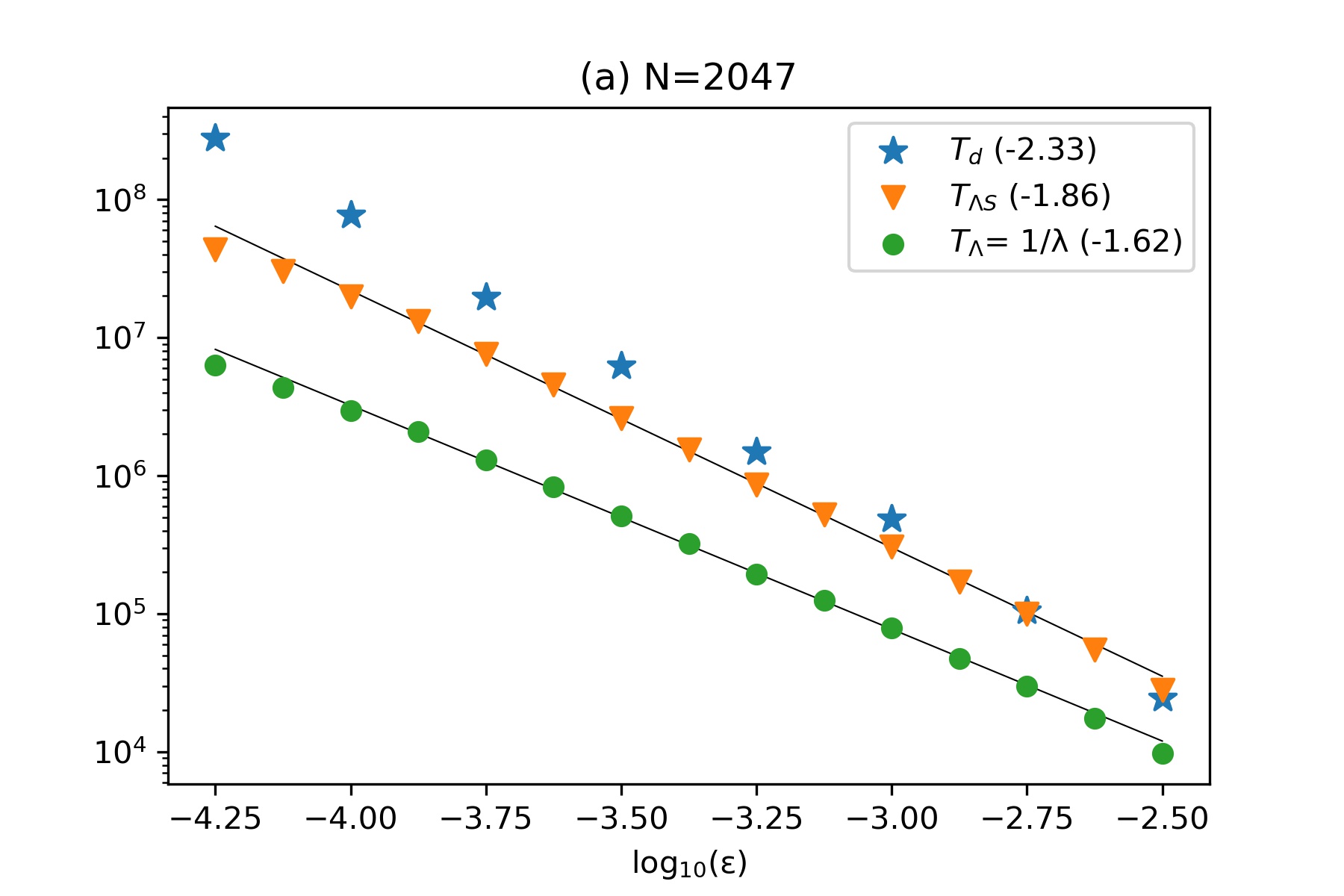}\\
      \includegraphics[width=0.75\linewidth]{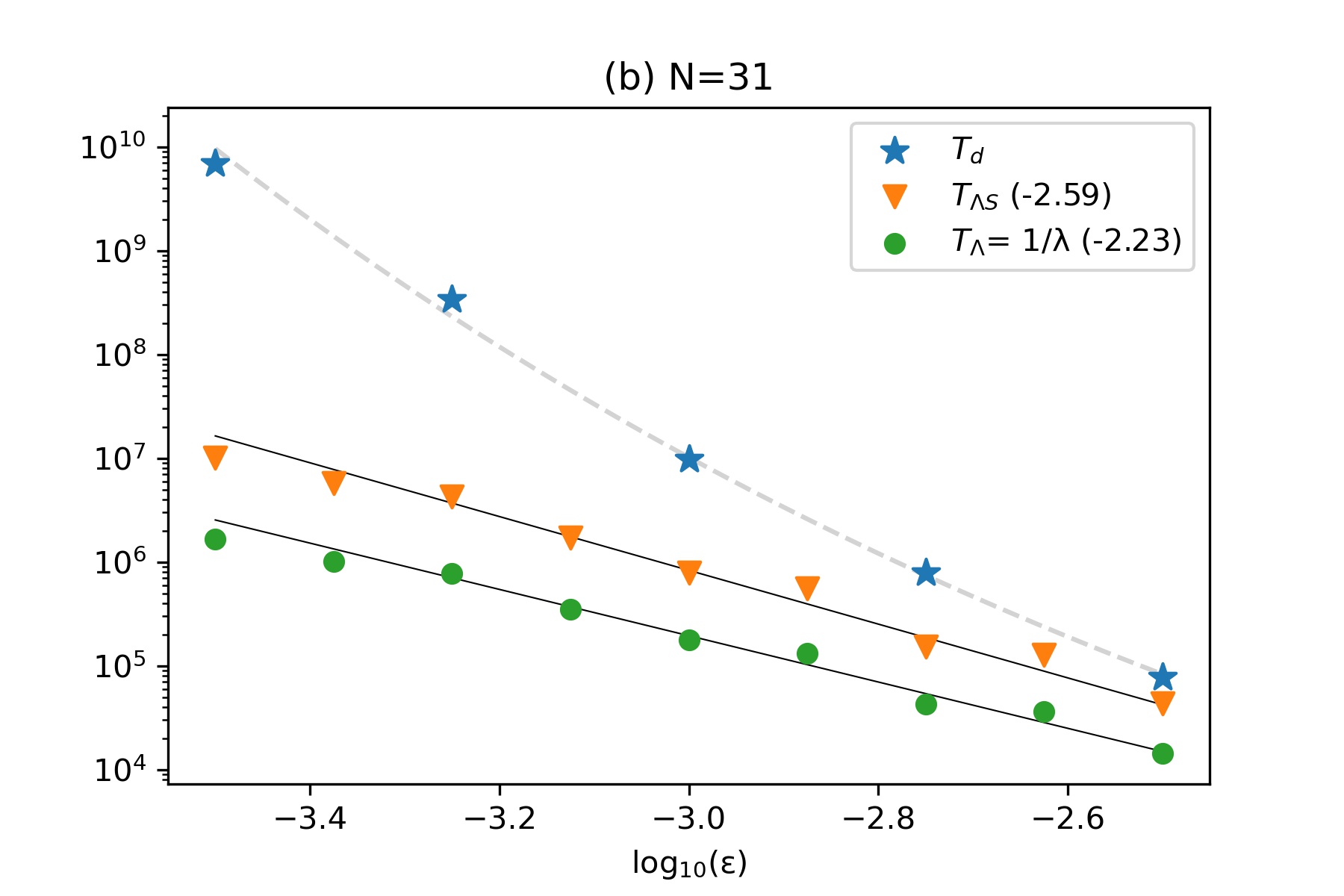}
    \caption{ Comparison of three timescales: $T_d$, $T_{\Lambda S}$ and $T_{\Lambda}$ versus the logarithm of the energy density, $\log_{10} \varepsilon$, for the FPUT system with (a) $N=2047$, and (b) $N=31$ particles. }
    \label{comparison}
\end{figure}

In Fig. \ref{All_LEs} is shown the dependence of the maximum Lyapunov exponent on the energy density $\varepsilon$, evaluated for  $N=2^n-1$, from $31$ to $2047$ particles.
As $N$ increases, the slope gradually converges to $1.6$. On the other hand, there seem to be two energy density regimes for $N=31$ and $63$ particles, characterized by distinct slopes. The last point for $N=31$ at $\varepsilon = 10^{-4.25}$ has not been included because $\lambda(t)$ did not converge until $t \approx 3 \times 10^{10}$, the endpoint of the simulation. 

In the literature, Lyapunov exponent scaling-laws for high $N$ are similar: 
In [\onlinecite{Casetti}]
as well as in [\onlinecite{Lichtenberg}]
the authors report $\frac{5}{3} \approx 1.6$ for the pure $\alpha$ model.
In [\onlinecite{BenLyap2018}], Benettin at al. measure that the slope is $1.53$ for the $\alpha + \beta$ FPUT model and $1.57$ for Toda with a large time-step, imitating a perturbed Toda model. More recently, Goldfriend \cite{Goldfriend} finds the slope $1.55$ for a randomly perturbed Toda model. 

We investigate the dependence of the  maximum Lyapunov exponent $\lambda$ and its saturation time $T_{\Lambda S}$ on the energy density to compare it with the timescales found in section \ref{chapter_diffusion}.  We also compute the Lyapunov time  $T_{\Lambda}=1/\lambda$, the characteristic time for a system to display chaos. 

In the exemplary case of Fig.\ref{comparison}(a) for $N=2047$ particles, we compare the timescales $T_{\Lambda}$, $T_d$, and $T_{\Lambda S}$ within  a range of energy densities.
We note that all Lyapunov-exponent-related results ($\lambda$, $T_{\Lambda}$ and $T_{\Lambda S}$) are based on a single random initial condition, while $T_d$ results have been averaged over 20 random initial conditions. By performing a best linear fit on the displayed data, we find that the Lyapunov exponent scales with the energy density as
\begin{eqnarray} \label{lameps}
    \lambda (\varepsilon) \approx 0.8 \cdot \varepsilon ^{1.6} ~~,
\end{eqnarray}
while its saturation time scales as
\begin{eqnarray} \label{lameps}
    T_{\Lambda S} (\varepsilon) \approx 0.78 \cdot \varepsilon ^{-1.86} ~~.
\end{eqnarray}
From the above two laws, it becomes evident that the Lyapunov exponent and its characteristic time are more sensitive than the diffusion time $T_d$. In Fig.
\ref{comparison}(a), despite the fact that $T_{\Lambda S}$ and $T_d$ are approximately 
equal at higher energies,  the gap between them widens in the low energy limit, reaching one order of magnitude at $\varepsilon = 10^{-4.25}$.
The time interval from $T_{\Lambda S}$ to $T_d$ explains the duration required for a weakly chaotic system to begin exhibiting energy diffusion.
In a case of Arnold diffusion, this gap should be exponentially long in terms of $1/ \varepsilon$. Panel (b), which is analogous to Fig.\ref{comparison}(a) but for $N=31$ particles, indicates prolonged stability intervals during which diffusion is absent, raising the question of whether this case indicates the onset of a KAM regime. 

\section{ Exponentially-long timescales}\label{final}
\begin{figure}[H]
    \centering
    \includegraphics[width=0.52\linewidth]{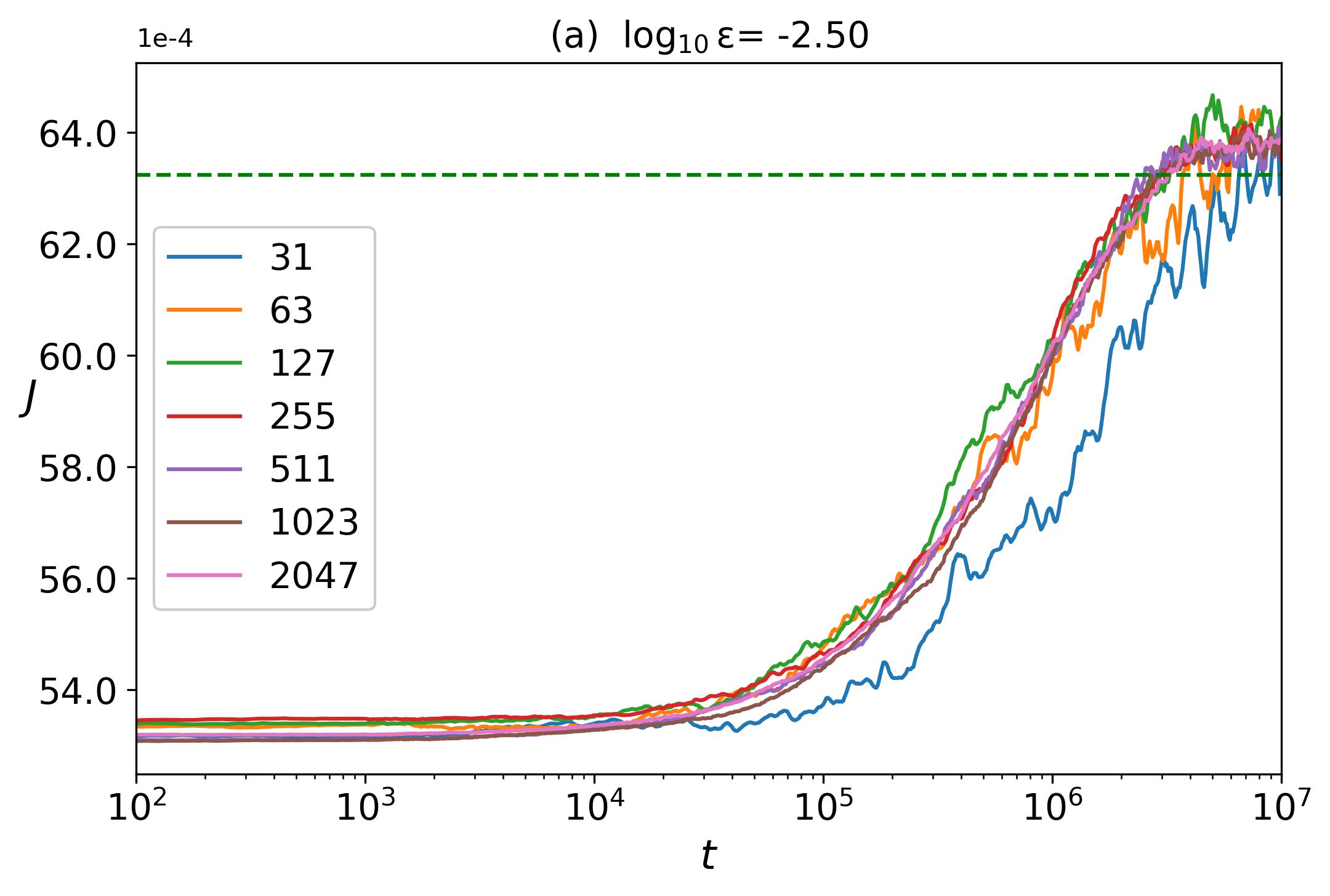}
    \includegraphics[width=0.5\linewidth]{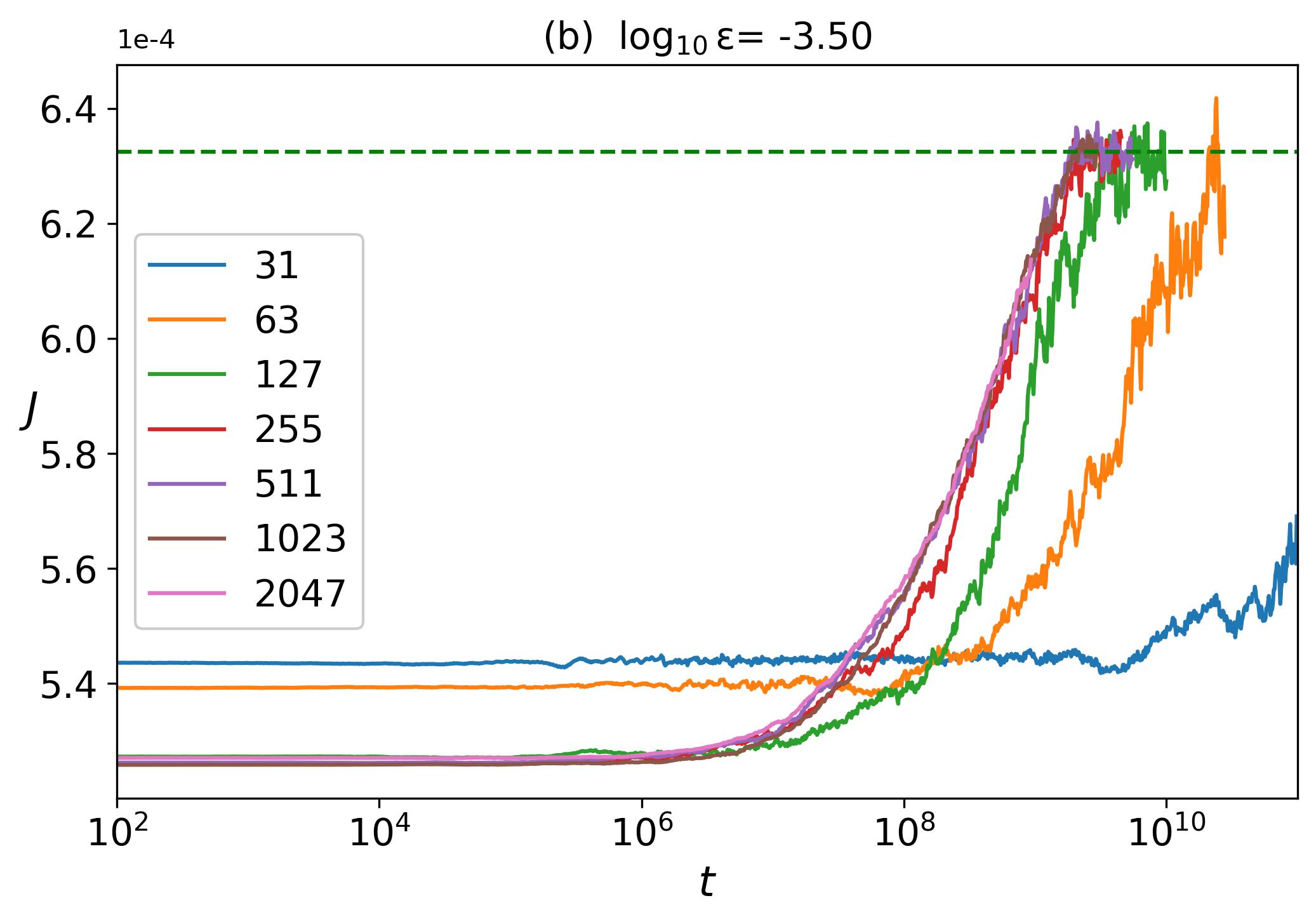}
    \includegraphics[width=0.53\linewidth]{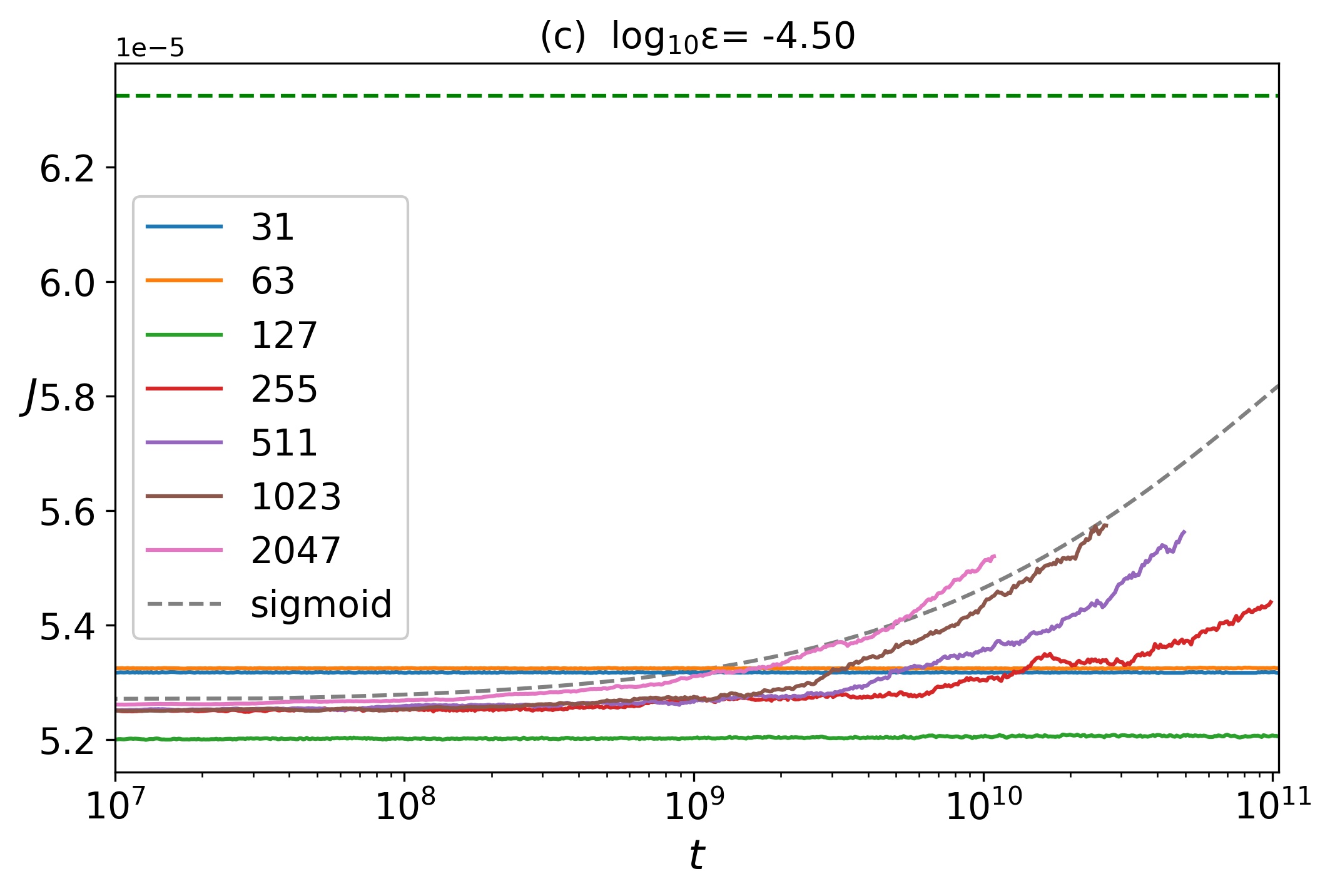}
    \caption{The temporal evolution of the Toda integral along FPUT dynamics for varying $N$ values that range from 31 to 2047. Panels correspond to the energies: (a) $\log_{10} \varepsilon = -2.5$, 
    (b) $\log_{10} \varepsilon = -3.5$. (c)  $\log_{10} \varepsilon = -4.5$. In panel (c) there are no signs of energy diffusion for $N=31,64,127$, indicating a KAM regime. The gray dashed line represents the sigmoid curve (\ref{sigmoid}) for the expected diffusion matching $N=2047, 1023$ case, and the gradual divergence for $N=511, 255$.  The top green dashed line represents $J_{eq} \approx 2 \varepsilon$. 
    }
    \label{All_J_plots}
\end{figure}

Fig. \ref{All_J_plots} displays the evolution of $J$ at three sample energies, 
where each curve is the average of 20 realizations. As the energy density decreases from panel (a) to panel (c), the timescales $T_d$ for $N \ge 2047$, which indicate phase-space diffusion, 
expand according to the law (\ref{td}) following $\varepsilon ^{-2.33}$.  

In this final section, we dive into very low energies to observe exponentially-long timescales, possibly linked to the onset of KAM-like regimes. 
 {The applicability of the KAM theorem in the FPUT problem is an open question, with some of the most notable contributions published by B. Rink  \cite{Rink2001,Rink2003}. In our work, we rely on numerical results that hint exponentially-long times to equilibrium, which are consistent with Nekhoroshev’s estimates \cite{Nekhoroshev1977, CaratiMaiocchi2012, HenriciKappeler2009}.}
We numerically examine the Toda integral $J$ from (\ref{j1}) at a fixed energy density for a range of system sizes of $N=2^n-1$, starting from $31$ and reaching $2047$ particles. 
We compare these results with the uniform sigmoids described in Eq.(\ref{sigmoid}) in section \ref{chapter_diffusion}.
At the lowest energies we were able to reach, computations were allowed to run until times longer than $10^{10}$. It is worth mentioning that all individual simulations exceeded $1260$ and required over two years to complete on the computer cluster.  

\begin{figure}[H]
    \centering
    \includegraphics[width=0.75\linewidth]{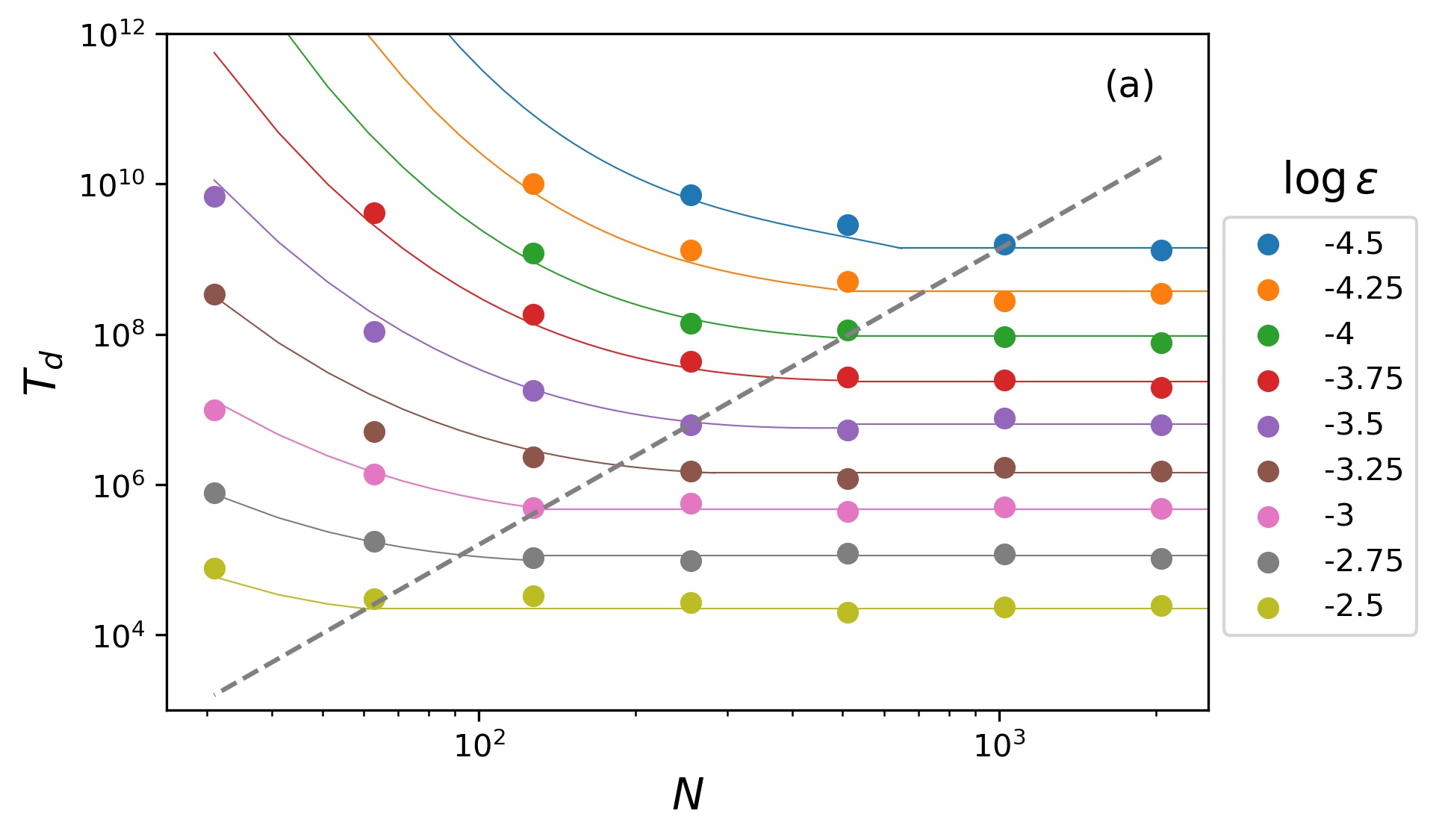}
    \includegraphics[width=0.75\linewidth]{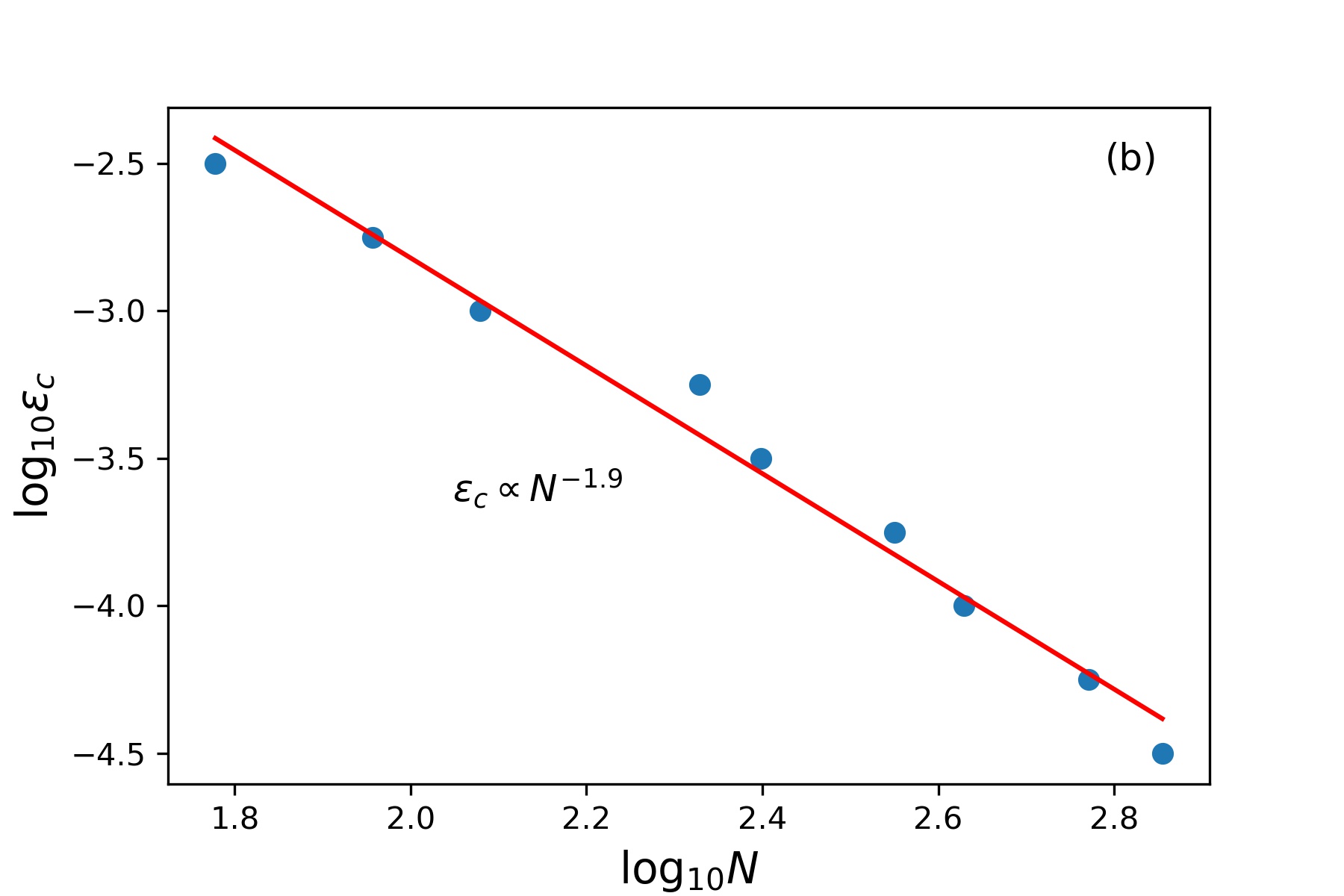}
    \caption{ (a) The onset-of-diffusion $T_d$ timescale for the Toda variable $J$ as a function of the system size $N$ at various energy levels, with a legend displayed on the right side of the panel. At each energy level, a crossover is observed separating the constant timescales from those which exponentially diverge with $1/N$. This borderline is marked by the dashed line.  
    (b) The energy density crossover $\varepsilon _c \propto N^{-1.9}$ corresponds to the dashed line in 
 panel (a) separating constant and divergent regimes of $T_d$. The data points correspond to the intersection of the two laws. The red line indicates that the energy density threshold $\varepsilon_c$ decreases  approximately as $N^{-2}$. }
    \label{TD}
\end{figure}

The Toda integral $J$ for lower $N$ sizes, however, diverges from the uniform $N$-independent sigmoid curve (\ref{sigmoid}) observed in larger systems. 
At $\varepsilon =10^{-2.5}$, Fig. \ref{All_J_plots}(a) shows an overall good agreement among the Toda integrals $J$ for different $N$ values. However, the diffusion for $N=31$ proceeds slightly more slowly compared to the other cases. In Fig. \ref{All_J_plots}(b), where the energy is $\varepsilon =10^{-3.5}$, $J$ for $N=127$ slightly disagrees with the curves for higher $N$. 
Here, $J$-timescales for $31$ and $63$ particles diverge significantly. At $\varepsilon =10^{-4.5}$, $N=255$ is evidently no longer part of the uniform
group of higher $N$ values, while $N=31$, $64$  and $127$ do not show any sign of energy diffusion. 
The latter three cases raise questions about whether $J$ is going to rise and reach $J_{eq}$, indicating a possible KAM regime. This numerical outcome for $N=127$ is in contrast to the theoretical predictions of a $1/N^{-4}$ energy density threshold \cite{Shepelyansky,Bambusi2015}, which yields $\varepsilon _c \approx 4 \times 10^{-9}$.


 We systematically collect and analyze the first timescale $T_d$ for all $N$ and energy densities $\varepsilon$. Fig.\ref{TD}(a) displays the plot of $T_d$ versus $N$ in double logarithmic scale. Each color represents a different energy density level. $T_d$ is independent of $N$ above a certain system size.  When the energy density $\varepsilon$ and $N$ values are sufficiently low, for example on the left side of the plot and above the dashed line, $T_d$ starts depending on $N$, increasing exponentially with $1/N$. For $\log_{10} \varepsilon \leq -3.5$, this exponential divergence is more evident, however, some points are  missing from the plot. These missing points corresponds to the completely flat $J$ cases, such as $N=31$, $64$  and $127$ lines in \ref{All_J_plots}(c), indicating a complete absence of diffusion where $T_d$ could not be determined. 

From Fig.\ref{TD}(a) we postulate the existence of a crossover between the energy density $\varepsilon$ and $N$, below which timescales are not uniform but increase exponentially with $1/N$, which is marked with the gray dashed line. 
In Fig. \ref{TD}(b) panel we plot this
energy crossover $\varepsilon _c$  versus $N$ in double logarithmic scale. We estimate that it decays with $N$ like
\begin{eqnarray}\label{e_c}
\varepsilon_c (N) = 11 \cdot N^{-1.9}  . 
\end{eqnarray}
The red line in  Fig. \ref{TD}(b), the linear fitting of the data,  corresponds to the dashed gray line of {Fig. \ref{TD}(a).

Remarkably, this decay in  $N$ has an exponent that is approximately the half of what is predicted from perturbation theory \cite{Shepelyansky,Bambusi2015}, where a KAM-like regime energy density threshold decays as $N^{-4}$.

\section{Conclusions}
In this work we measured thermalization of the FPUT system.
In particular, we employed a constant of motion of the integrable Toda lattice, denoted by $J$ and described in Eq.(\ref{j1}), which is an adiabatic invariant for the FPUT model. 

Our findings show that this quantity serves as an important tool which can detect, measure and model diffusion in the phase space of the FPUT system. 
In particular, the truncated Taylor expansion of $J$ to quadratic terms, see Eq.(\ref{quad}), is related to  
the covariance of momenta $\mathbf{p}$ and the covariance of relative displacements $\mathbf{\delta q}$, 
linking {\it decorrelation times}, at which the covariance vanishes, to {\it equilibrium times} $T_{eq}$.
 The sigmoid law of Eq.(\ref{sigmoid}) provides a theoretical description of the diffusion processes in FPUT. It models all stages of diffusion, estimating two essential timescales: $T_d  \propto \varepsilon ^{-2.33}$, the onset of diffusion and, $T_{eq}  \propto \varepsilon ^{-2.75}$, the equilibrium time. 
 These timescales grow algebraically with lowering the energy density and are uniform, independent of the number of particles $N$, as long as the system size $N$ is large enough. However, at smaller system sizes we find a critical energy density $\varepsilon_c \propto N^{-2}$ below which thermalization is substantially slowing down.  
The latter result is crucial to the FPUT problem: The critical energy $\varepsilon_c$ could indicate the onset of a KAM regime, which vanishes in the thermodynamic limit with rate $1/N^2$, instead of $1/N^4$ as predicted from some perturbation theory approaches \cite{Shepelyansky,Bambusi2015}.
Our numerical results suggest that FPUT is ergodic in the thermodynamic limit. 
It remains open to find a way to derive the critical energy density, $\varepsilon_c \propto 1/N^2$, might have to include the proximity of FPUT to the Toda lattice.

To complete the current study, we examined the maximum Lyapunov exponent $\lambda$ and its relevant timescales $T_{\Lambda }=1/\lambda$ (Lyapunov characteristic time), as well as the Lyapunov saturation timescale $T_{\Lambda S}$, comparing them with $T_d$ and $T_{eq}$.
While $T_{\Lambda S}$ and $T_d$ are nearly identical at higher energies, in the low-energy limit, the FPUT system displays longer stability intervals without the presence of diffusion. 
We conclude that 
the Lyapunov exponent and its characteristic times ($T_{\Lambda}$, $T_{\Lambda S}$) are the first to detect weakly chaotic behavior, hence are more sensitive than the adiabatic invariant $J$. 

\section{Acknowledgments}
SF supported by the Institute for Basic Science through Project Code (No. IBSR024-
D1) and HC by the EPSRC New Investigator Award UKRI3307.
\section{Appendix: Statistical Analysis and curve fitting}
\subsection{Energy spectrum and standard deviation}
 {This section includes additional figures related to Fig.\ref{paradigm}.}
\begin{figure}[H]
\centering
    \includegraphics[width=0.7\linewidth]{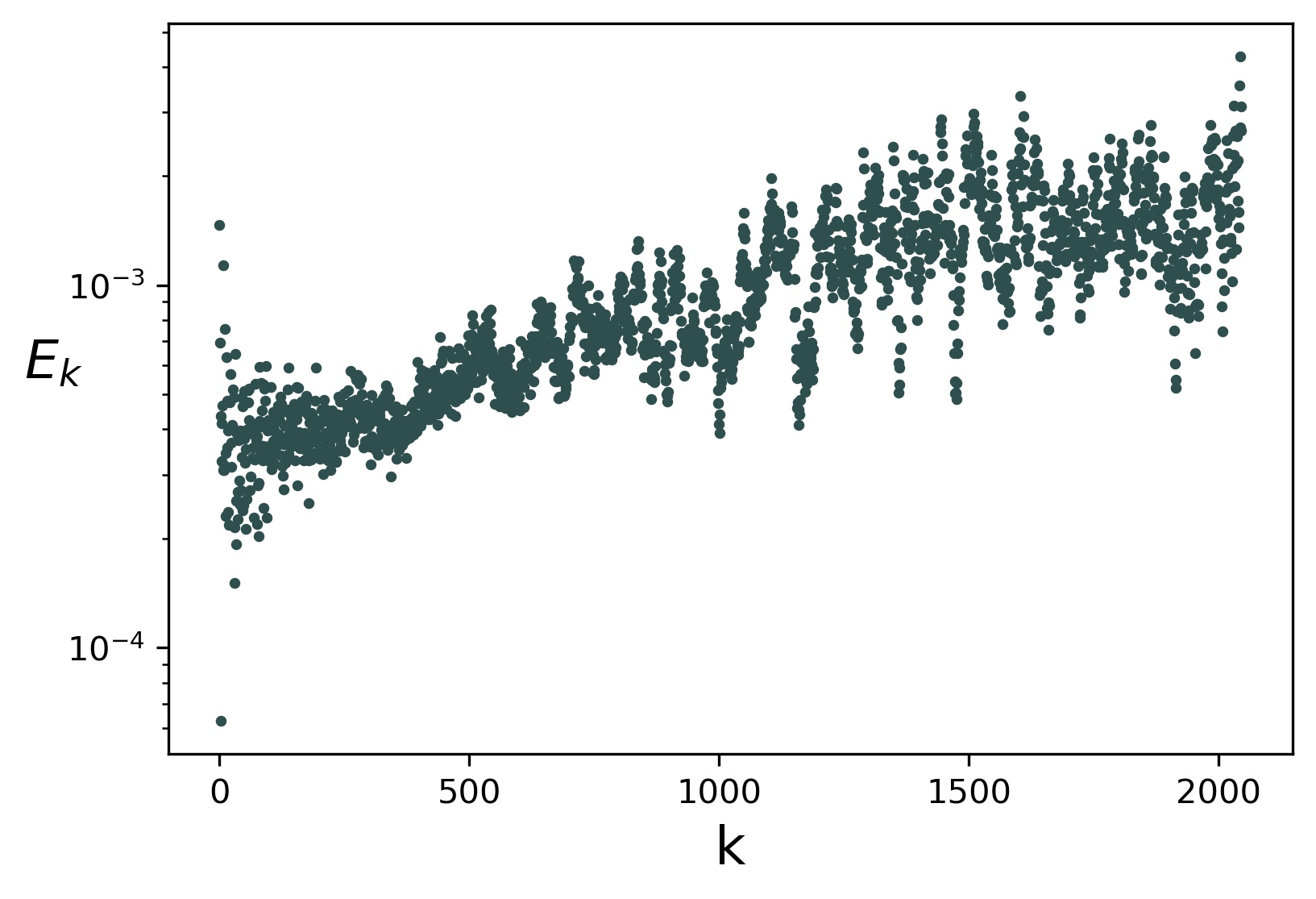}
    \caption{ The time-averaged over $[0, 10^5]$ energy spectrum of FPUT with $N=2047$ particles at an energy level $\varepsilon = 10^{-3}$. }
    \label{Eks}
\end{figure}
\begin{figure}[H]
    \centering
\includegraphics[width=0.63\linewidth]{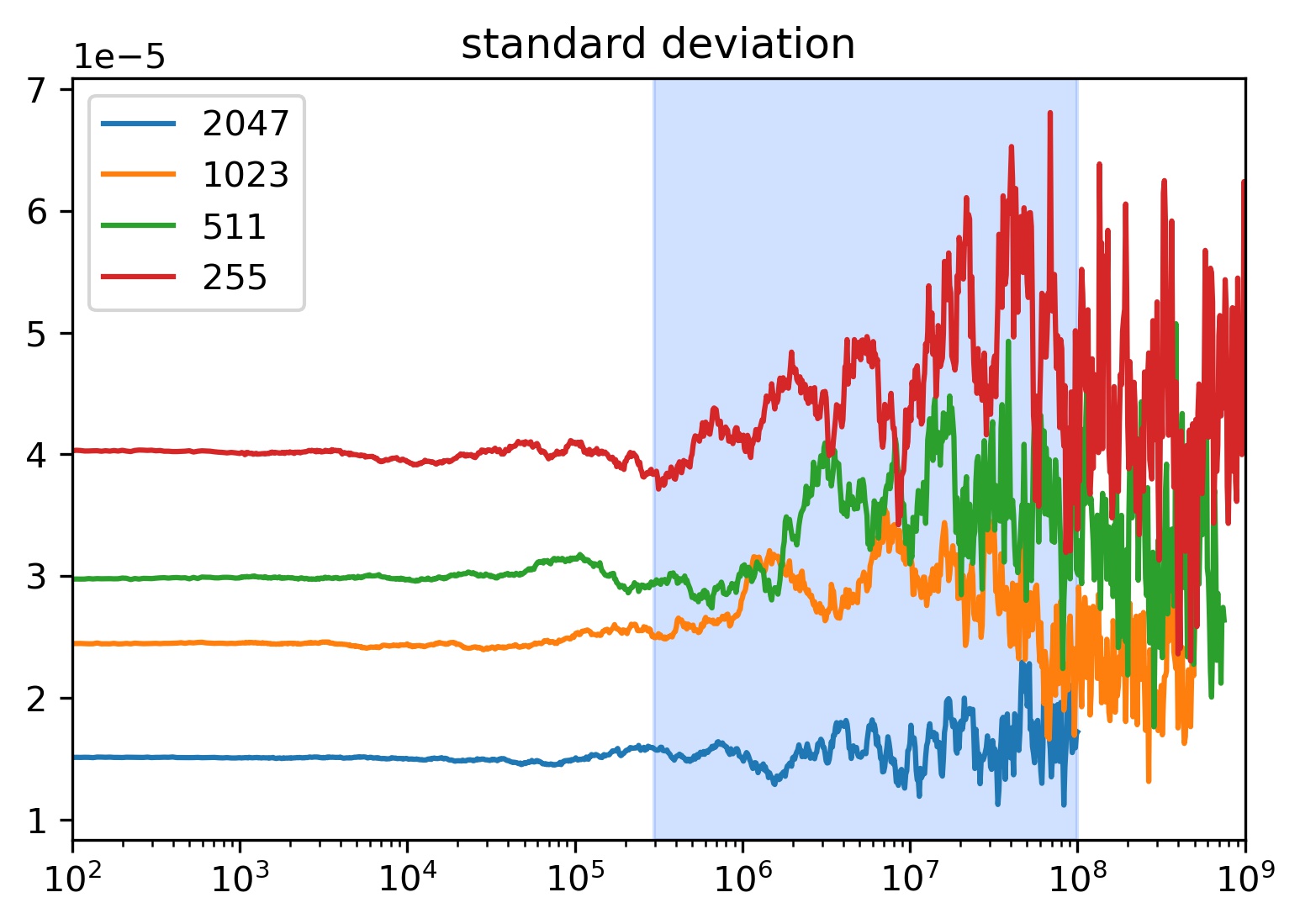}
    \caption{
The evolution of the standard deviation from 20 Toda integrals along FPUT for 4 different $N$ values. 
The light-blue area highlights the energy diffusion window from Fig.\ref{paradigm}.}
    \label{std1}
\end{figure}

\subsection{ Sigmoid fittings \label{AppSigm}}
We fit the numerical data with the law Eq.(\ref{sigmoid}) for the system with $N=2047$ particles at various energies,  {as shown in Fig.\ref{sigmoid_fittings}. }
 We then determine the optimal $\gamma$ values 
  {by a best-curve fit algorithm.}
 At energies $\varepsilon \geq 10^{-3}$ it is  $\gamma \approx 1$, while at $\varepsilon < 10^{-3}$ becomes   $\gamma(\varepsilon)  \approx 2.9\varepsilon ^{ 0.15 }$  {(displayed in Fig.\ref{gamma_fit} with more details)}. To obtain an accurate  value for the equilibrium time $T_{eq}$,  { we set the Gibbs value $J_{eq}$ in Eq.(\ref{sigmoid}) to be $3\%$ larger than the expected $2\varepsilon$. This allows to allocate a finite value of $T_{eq}$, since the sigmoid curve in Eq.(\ref{sigmoid}) approaches $J_{eq}$ only asymptotically.}
     \begin{figure}
    \centering
    \includegraphics[width=0.485\linewidth]{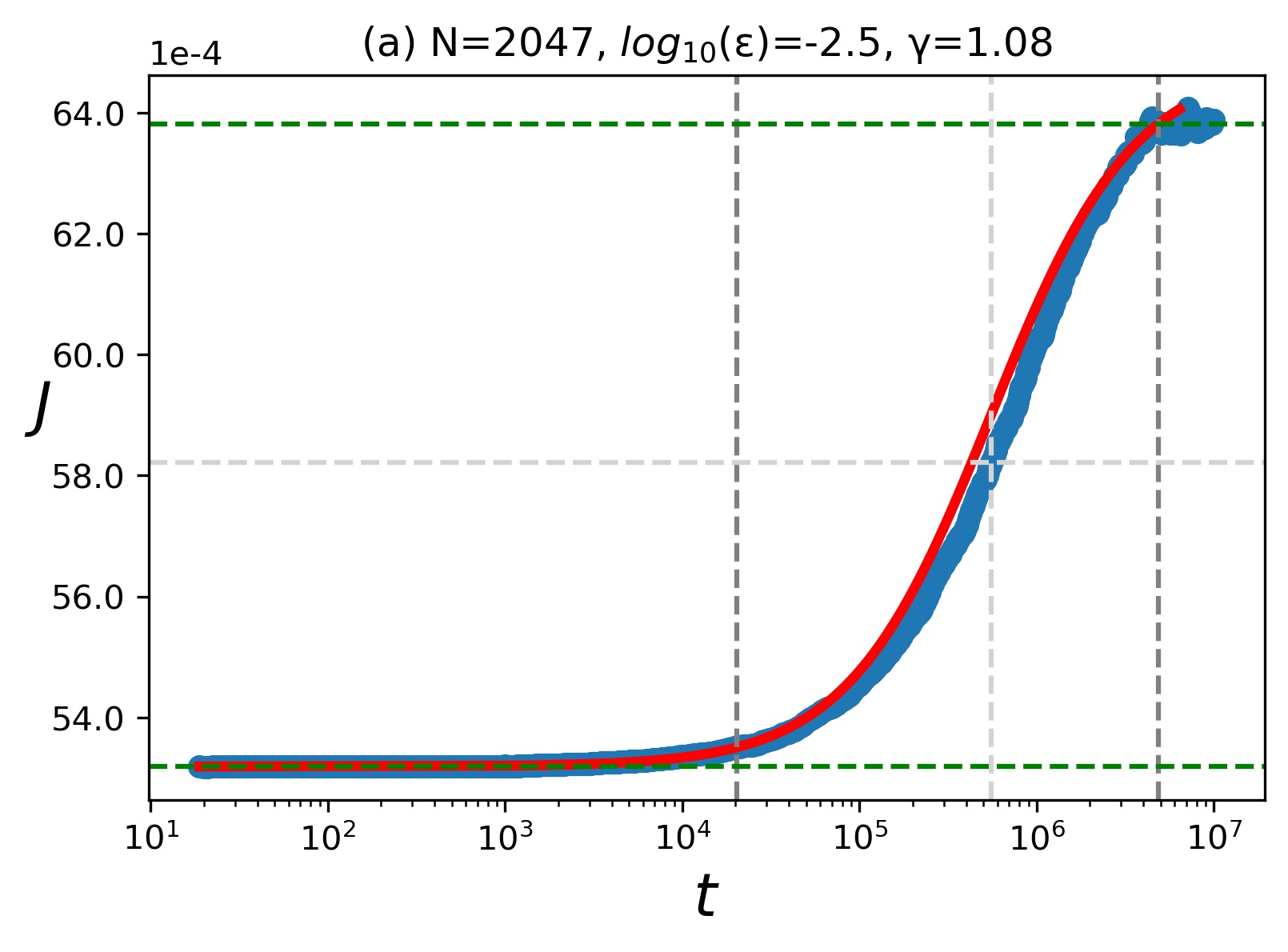}
    \includegraphics[width=0.485\linewidth]{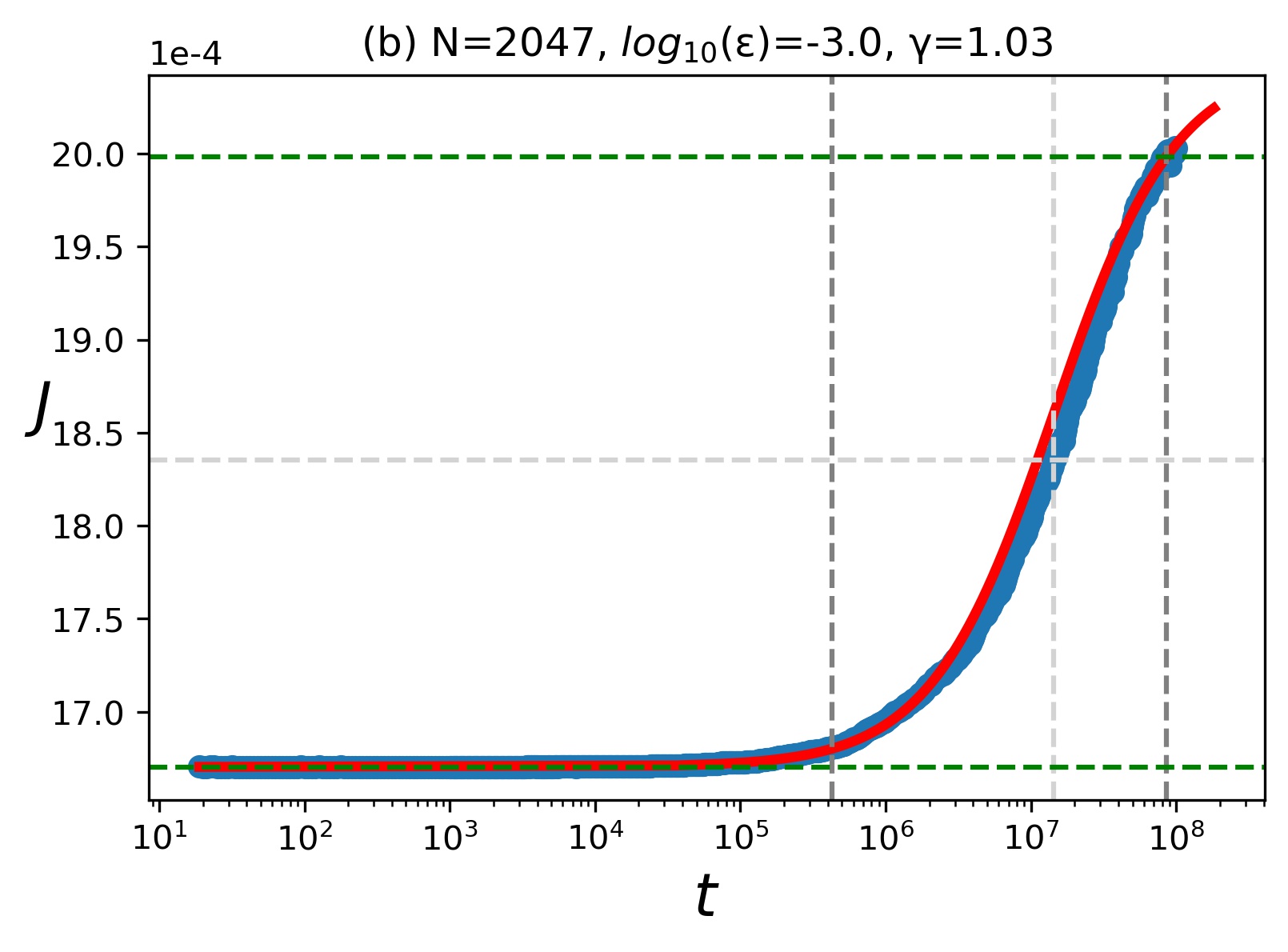}
    \includegraphics[width=0.485\linewidth]{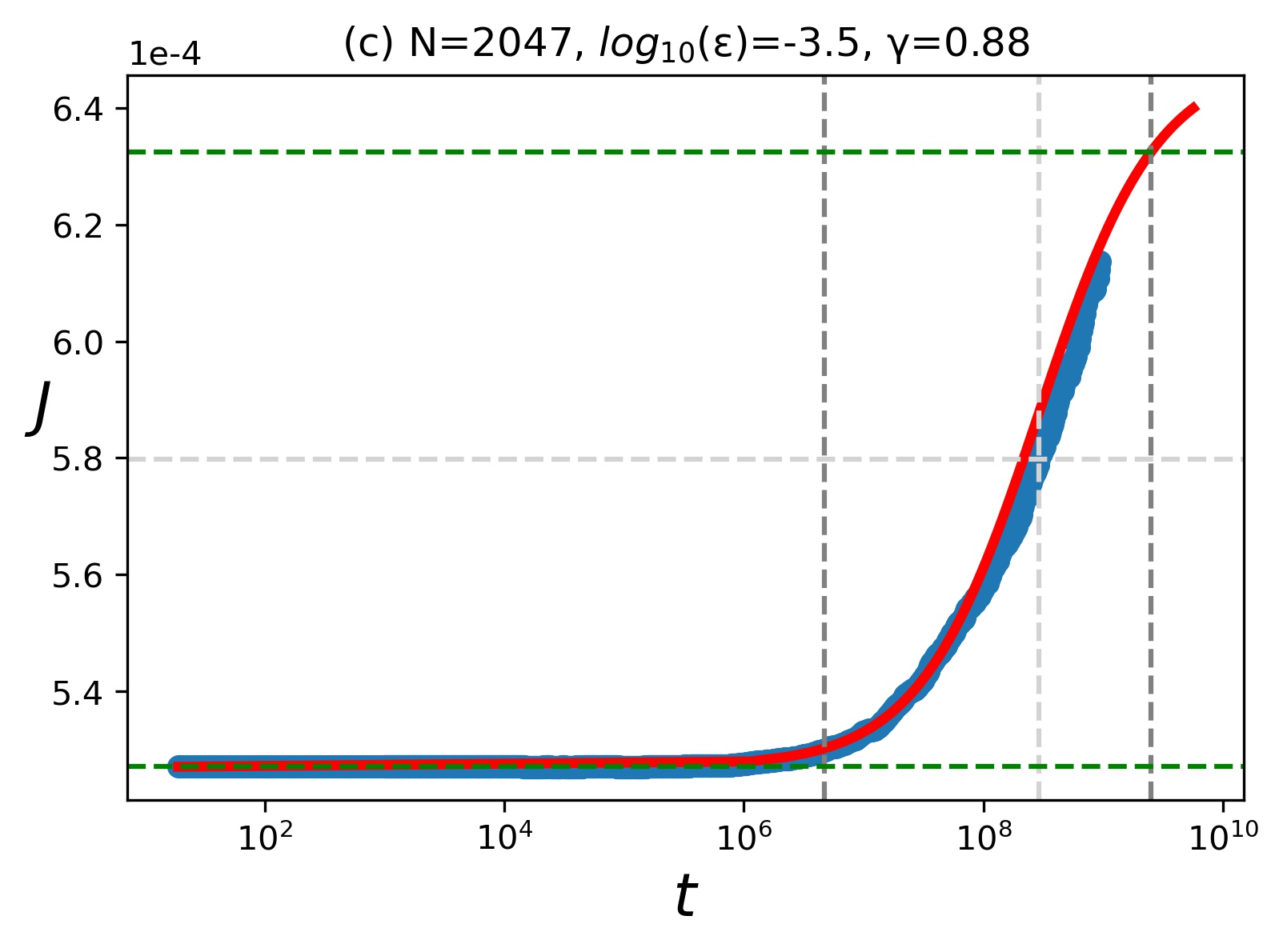}
     \includegraphics[width=0.485\linewidth]{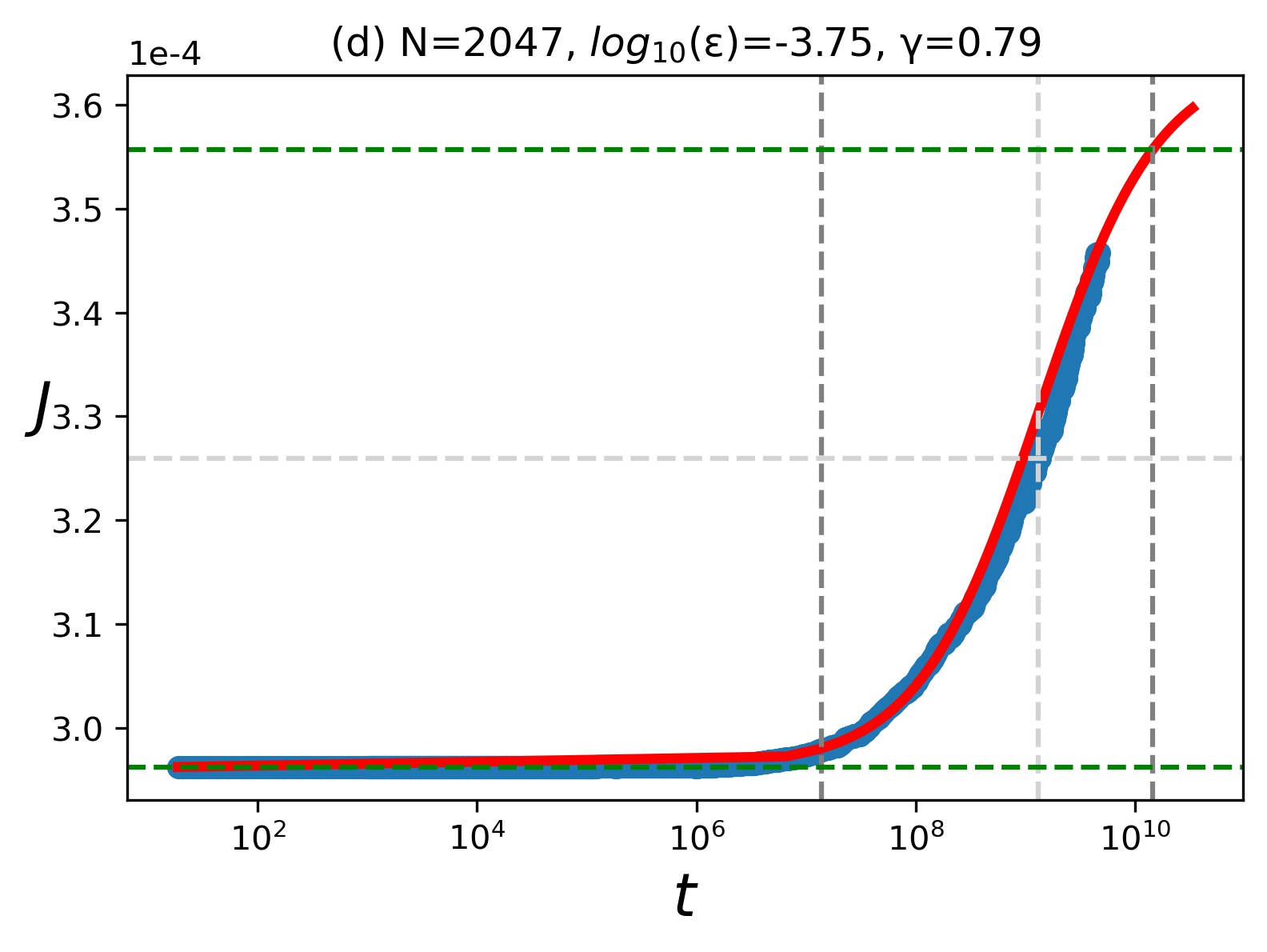} 
    \caption{ The evolution of $J$ along FPUT for $N=2047$ at various energy density values (blue) and the sigmoid best fittings of Eq.(\ref{sigmoid}) with optimal  exponent $\gamma=\gamma(\varepsilon)$ values (red curve). The left and right vertical (grey) dashed lines correspond to the theoretical timescales $T_d$ and $T_{eq}$, derived from Eq.(\ref{sigmoid}). The (light grey) vertical dashed line in the middle is for $t_0$. The (green) top and bottom horizontal dashed lines denote $J_0$ and $J_{eq}$, respectively. 
    }
    \label{sigmoid_fittings}
\end{figure}

     \begin{figure}
    \centering
    \includegraphics[width=0.75\linewidth]{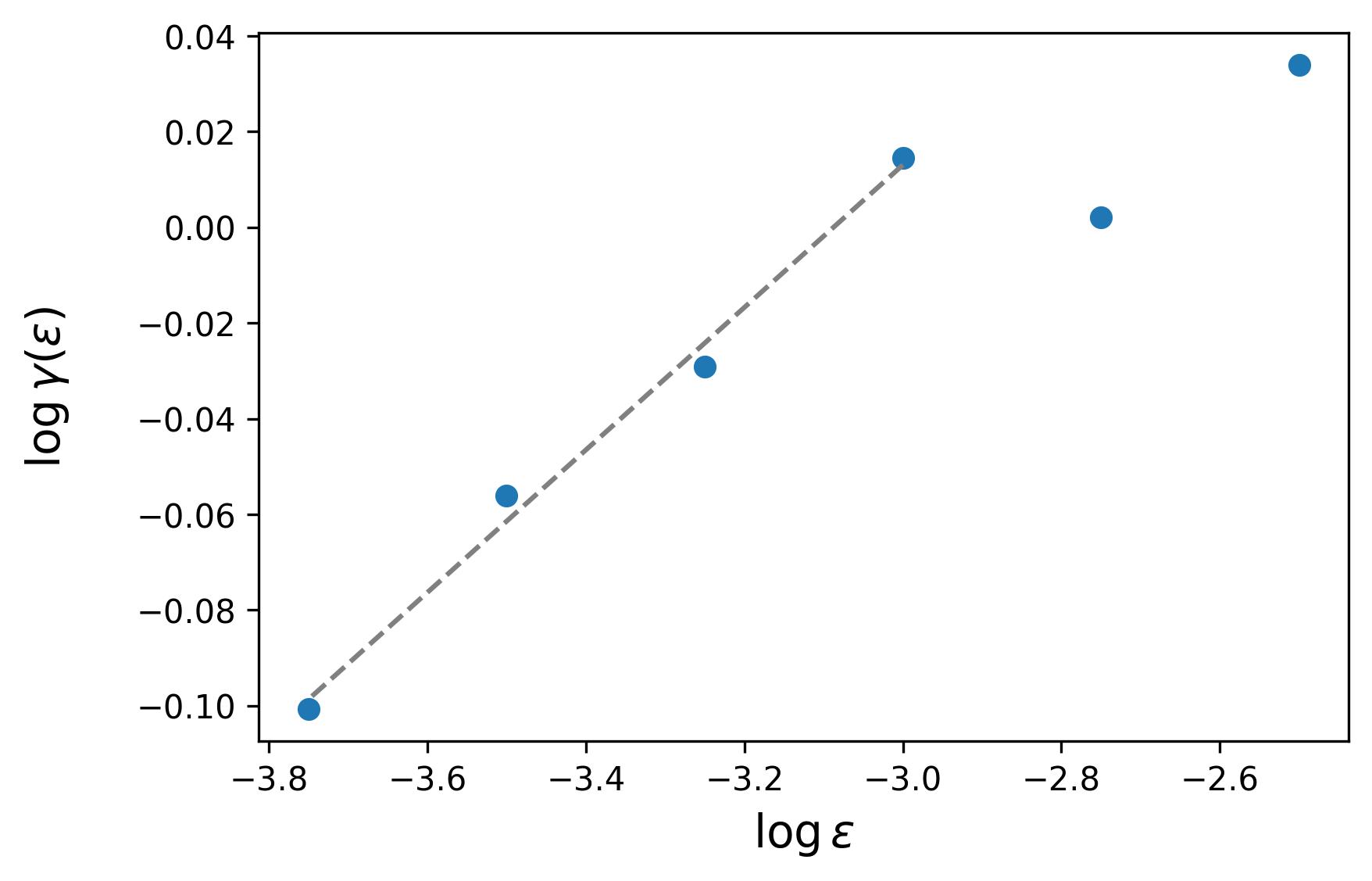}
 \caption{  { The fitted $\gamma$ values, which optimize the sigmoid fit of Eq.(\ref{sigmoid}) to the numerical data for $J$, are plotted versus the logarithm energy density $\varepsilon$. They show a linear increase for energy values lower than $10^{-3}$, highlighted by the dashed gray line which represents the power-law  
 $\gamma =  2.88 \varepsilon ^{0.148}$.}}
      \label{gamma_fit}
\end{figure}


\bibliography{references,sergejflach}

\end{document}